\pdfoutput=1
\documentclass[12pt]{article}

\usepackage{graphicx,psfrag,epsf,color}
\usepackage{amsmath,amssymb,amsfonts}
\usepackage{hyperref}
\usepackage{array}
\usepackage{bm} 
\usepackage{cite}
\usepackage{hyperref}
\usepackage{multirow}
\usepackage{rotating}
\usepackage{slashed}
\usepackage{textcomp}
\usepackage{cancel}
\usepackage{arydshln}

\newcounter{MBQ}

\newcounter{HTQ}

\newcommand{\lessim}{\mbox{\raisebox{-3pt}{$\stackrel{<}{\sim}$}}}

\newcommand{\be}{\begin{equation}}
\newcommand{\ee}{\end{equation}}
\newcommand{\bea}{\begin{eqnarray}}
\newcommand{\eea}{\end{eqnarray}}
\newcommand{\bi}{\begin{itemize}}
\newcommand{\ei}{\end{itemize}}
\newcommand{\ben}{\begin{enumerate}}
\newcommand{\een}{\end{enumerate}}
\newcommand{\bt}{\begin{tabular}}
\newcommand{\et}{\end{tabular}}

\newcommand{\lp}{\left(}
\newcommand{\rp}{\right)}
\newcommand{\LQCD}{\Lambda_{\rm QCD}}

\newcommand{\mtau}{m_\tau}

\newcommand{\LMS}{\Lambda_{\overline{\rm MS}}}
\newcommand{\non}{\nonumber \\}

\numberwithin{equation}{section}
\allowdisplaybreaks

\begin{document}
\allowdisplaybreaks

\begin{titlepage}

\begin{flushright}
{\small
TUM-HEP-1575/25\\
YITP-25-162\\
13 October 2025
}
\end{flushright}

\vskip1cm
\begin{center}
{\Large \bf 
Gradient-flowed operator product expansion without IR renormalons
\\[0.2cm]}
\end{center}

\vspace{0.45cm}
\begin{center}
{\sc M.~Beneke$^{a}$}
and  {\sc H.~Takaura$^{b}$}\\[6mm]
{\it ${}^a$Physik Department T31,\\
James-Franck-Stra\ss e~1, 
Technische Universit\"at M\"unchen,\\
D--85748 Garching, Germany}
\\[0.3cm]
{\it ${}^b$Center for Gravitational Physics and Quantum Information,\\ Yukawa Institute for Theoretical Physics, Kyoto University,\\
Kyoto 606-8502, Japan}
\\[0.3cm]
\end{center}

\vspace{0.55cm}
\begin{abstract}
\vskip0.2cm\noindent 
A long-standing problem concerns the question how to consistently combine perturbative expansions in QCD with power corrections in the context of the operator product expansion (OPE), since the former exhibit ambiguities due to infrared renormalons, which are of the same order as the power corrections. We propose to use the gradient flow time $1/\sqrt{t}$ as a factorization scale and to express the OPE in terms of IR renormalon-free subtracted perturbative expansions and unambiguous matrix elements of gradient-flow regularized local operators. We show on the example of the Adler function and its leading power correction from the gluon condensate that this method dramatically improves the convergence of the perturbative expansion. We employ lattice data on the action density to estimate the gradient-flowed gluon condensate, and obtain the Adler function with non-perturbative accuracy and significantly reduced theoretical uncertainty, enlarging the predictivity at low $Q^2$. When applied to the hadronic decay width of the tau lepton, the method resolves the long-standing discrepancy between the fixed-order and contour-improved approach in favour of the fixed-order treatment.
\end{abstract}
\end{titlepage}



\section{Introduction}
\label{sec:introduction}

The operator product expansion (OPE) is a fundamental framework to calculate the short-distance behaviour of products $A(x)B(0)$ of local operators \cite{Wilson:1972ee,Zimmermann:1972tv}. In QCD it gained importance in the context of the SVZ sum rules \cite{Shifman:1978bx, Shifman:1978by} and inclusive heavy-quark decays. In these applications the OPE factorizes the 
weak-coupling physics  at the short-distance scale in the correlation function of quark currents from the strong-coupling physics at the QCD scale $\Lambda$.  The OPE of the time-ordered two-point correlation function in Fourier-momentum space can be written in the form 
\be
\Pi(Q^2)= 
i\int d^4 x \,e^{i q x} \,\langle \,\mbox{T}\big[A(x) B(0)\big]\,\rangle = \sum_{n} 
\frac{C_{O_n}\!(Q^2)}{(Q^2)^{d_{O_n}/2}}\,\langle O_n \rangle,
\ee
where $Q^2=-q^2$ and the power in $Q^2$ of each term is determined 
by the mass dimension $d_{O_n}$ of the local operator $O_n$.\footnote{And the mass dimension of $\Pi(Q^2)$. In the above equation we assumed that $\Pi$ has been 
defined such that it is dimensionless. In the case of heavy quarks, $Q$ can be the heavy-quark mass, and the expectation values $\langle ...\rangle$ refer to averages in the heavy-hadron rather than the QCD vacuum state.}
The short-distance  coefficients $C_{O_n}\!(Q^2)$ are expanded in 
a perturbative series in the strong coupling $\alpha_s(Q)$,
whereas the matrix elements $\langle O_n \rangle$ encode 
the information concerning non-perturbative dynamics.

The method of choice to calculate the short-distance coefficients in terms of QCD loop diagrams is dimensional regularization, which defines the separation into small and large momenta via analytic continuation rather than a dimensionful cut-off. Each term of the OPE is then generally ill-defined: the short-distance coefficients develop a non-sign-alternating factorial divergence known as infrared (IR) renormalons, 
since dimensional regularization removes only the logarithmically divergent IR 
behaviour. The power divergences of local operators, which are set to zero 
in dimensional regularization, in turn show up as ambiguities of the operator matrix elements after analytic continuation to four dimensions.\footnote{See ref.~\cite{Beneke:1998ui} for a review on this topic and an extensive list of original references.} These features originally triggered a debate on the meaning and validity of the OPE in QCD \cite{David:1983gz, David:1985xj,Novikov:1984rf}, but it is now understood that the OPE viewed as a double expansion in 
$\alpha_s(Q)$ and the exponentially small (in $\alpha_s(Q)$)  quantity $\Lambda/Q$  
fits the modern view of a trans-series interpretation, which can be verified in 
the $1/N$ expansion of the two-dimensional $O(N)$ non-linear sigma model 
\cite{David:1983gz,Beneke:1998eq,Liu:2025bqq}. The ambiguities in summing non-sign-alternating divergent series and in the operator matrix elements are therefore 
expected to cancel when adding corresponding terms in the OPE. 
Nevertheless, the problem remains how one can systematically improve 
the precision and consistently include power corrections in the OPE, 
as none of the all-order perturbative series and none of the 
corresponding all-order subtractions of power-divergent non-perturbative 
matrix elements can be computed in practice.

The IR renormalon problem is not merely a formal issue but is becoming increasingly important from a phenomenological viewpoint, because 
modern loop integral methods allow for the calculation of orders high enough that issues related to factorial divergence limit the theoretical accuracy.  For instance, two seemingly reasonable perturbative approximations to the hadronic tau lepton decay width do not agree with each other and the difference has been attributed to the renormalon 
problem \cite{Beneke:2008ad} and an inconsistency with the operator product expansion \cite{Gracia:2023qdy} in one of the two approaches. The factorial divergence due to an IR renormalon is even more prominent for the pole mass of a heavy quark \cite{Beneke:1994sw,Bigi:1994em} and requires subtractions. While in this case, the problem is often spurious and can be solved by employing another mass definition \cite{Beneke:2021lkq}, in which the leading IR renormalon is absent, the two-point correlations functions above provide access to a wealth of physical quantities in hadronic and heavy quark physics, which are probed through the respective external currents. It is then essential to overcome the IR renormalon problem within the OPE framework, if one aims at accuracy that includes power-suppressed terms.

Various renormalon subtraction methods have been proposed in recent years beyond the scope of the pole-mass renormalon issue \cite{Ayala:2019uaw,Hayashi:2021vdq,Hayashi:2023fgl,Benitez-Rathgeb:2022yqb,Benitez-Rathgeb:2022hfj}. The divergence pattern caused by IR renormalons can be parameterized by perturbatively calculable coefficients related to renormalization-group functions, and a number of overall constants (the Stokes constants). 
The methods of refs.~\cite{Hayashi:2021vdq,Hayashi:2023fgl} require these coefficients, and those of refs.~\cite{Ayala:2019uaw,Benitez-Rathgeb:2022yqb,Benitez-Rathgeb:2022hfj}
further need the Stokes constants. Obtaining the Stokes constants from low-order perturbative data by fitting the IR renormalon divergence pattern is delicate. 
Beyond the dominant renormalon, the renormalon structure is difficult to elucidate  because of more complicated mixing between different OPE terms of the same 
dimension. For this reason, the application of the above methods has rarely been attempted \cite{Hayashi:2021vdq,Hayashi:2023fgl,Takaura:2018vcy,Ayala:2020odx} beyond the simplest cases where there is only a single dominant renormalon and Stokes constant. 
Moreover, these methods do not provide by themselves a non-perturbative definition of the operator matrix elements.

In this paper, we propose a new method to achieve non-perturbative precision and to overcome the renormalon problem.\footnote{The basic idea and an application to the hadronic $\tau$ decay, which resolves the above mentioned problem, has been presented in ref.~\cite{Beneke:2023wkq}.} The central idea consists of 
applying the gradient flow \cite{Luscher:2010iy,Luscher:2011bx} to rewrite
the ambiguous and technically undefined matrix elements $\langle O_n \rangle$ 
in terms of well-defined matrix elements 
$\langle \mathcal{O}_n(t) \rangle$ at finite flow time $t>0$, where $1/\sqrt{2 t}$ acts effectively as a Wilsonian or hard ultraviolet (UV) cut-off, which defines 
the renormalized local operators. This induces a modification in the short-distance 
coefficients of operators of dimension smaller than $d_n$, in which the IR 
renormalons are cancelled. The so-defined {\em gradient-flowed OPE} can be considered as a realization of a ``Wilsonian" RGE with a hard momentum cut-off with the crucial advantages that gauge-invariance is maintained, operator matrix elements are suitable for lattice computations, and dimensional regularization and $\overline{\rm MS}$ renormalization can be employed for the short-distance coefficients as usual. The new OPE is not only free from the IR renormalon problem but also achieves non-perturbative precision, since 
power corrections are unambiguously defined and can be consistently combined with IR renormalon-free perturbative series.\footnote{After the publication of ref.~\cite{Beneke:2023wkq} we became aware of refs.~\cite{Monahan:2014tea, Monahan:2015lha}, 
which defines a ``smeared OPE" in a scalar field theory, which is technically identical to the gradient-flowed OPE constructed below for QCD correlation functions. The conceptual context is, however, very different. The authors of refs.~\cite{Monahan:2014tea, Monahan:2015lha} have in mind applications to the renormalization of moments of light-cone operators at leading power. The main result of the present work is that the gradient-flowed OPE provides a practical solution to the long-standing problem of how to combine IR renormalon-divergent series with non-perturbative power corrections.} 

Compared to previous methods of renormalon subtraction mentioned above, the present proposal offers some distinct advantages. Besides defining the local operators in a way that is ready for lattice computations, no information on the IR renormalon structure, in particular the Stokes constants, is required. The rearranged OPE subtracts the fixed-order coefficients automatically, assuming consistency of the standard OPE. By defining the gradient-flowed operators to higher order in the OPE series, gradient-flow subtraction can in principle deal with all IR renormalons, not only the dominant one.

In this paper we illustrate the gradient-flow subtraction method for the so-called Adler function, derived from the correlation function of quark vector-currents. The outline is as follows: In sec.~\ref{sec:gradienflowedOPE} we formulate the gradient-flow subtraction for the OPE of the Adler function including the leading power correction from the gluon condensate. The cancellation of the associated IR renormalon singularity in the perturbative Adler function is checked analytically in the so-called large-$\beta_0$ approximation. In the subsequent sec.~\ref{sec:adler}, we study the effectiveness and precision of the gradient-flowed OPE. We translate available lattice data into the gradient-flowed gluon condensate and identify an optimal window in gradient-flow time 
for the subtraction. We then provide non-perturbative results for the Adler function as a function of $Q$. Sec.~\ref{sec:tau} presents a discussion of the hadronic tau decay series similar to but improving upon ref.~\cite{Beneke:2023wkq}. 
An extension of the subtraction method to dimension-six operators in the OPE is 
investigated in sec.~\ref{sec:dim6} in a highly simplified setting.
We conclude in sec.~\ref{sec:conclusion}. Additional details are summarized in the Appendices. 


\section{Renormalon subtraction and gradient-flowed OPE of the Adler function}
\label{sec:gradienflowedOPE}

\subsection{Adler function}

We consider the two-point function 
\be
3 Q^2 \Pi(Q^2) = i \int d^4 x \, e^{i q x} \langle 0| T \big[J^{\mu}(x) J^{\dagger}_{\mu}(0)\big] | 0 \rangle, \qquad Q^2=-q^2
\label{eq:PI}
\ee
of the flavour-non-singlet vector current $j^\mu = \bar{u}\gamma^\mu d$ in QCD with $n_f=3$ massless quarks (up, down and strange). This set-up is the relevant one in the $Q=1\ldots 2\,$ GeV range, which is of particular interest for the OPE. The Adler function is defined as 
\be 
D(Q^2)=-Q^2 \frac{d \Pi(Q^2)}{dQ^2}
\ee
and depends only on the dimensionless ratio $Q/\Lambda$ or, alternatively, the strong coupling $\alpha_s(Q)$. 
We define $\alpha_s$ in the $\overline{\rm MS}$ scheme. 
The beta function, which governs the scale-dependence 
of $\alpha_s$, is defined as
\be
\beta(\alpha_s)=\mu^2 \frac{d \alpha_s}{d \mu^2}=\sum_{k=0}^{\infty} \beta_k \alpha_s^{k+2}\,. \label{eq:betafn}
\ee
With this convention the expansion starts with
$\beta_0=-(11N_c/3-2n_f/3)/(4 \pi)$. Here and below $\alpha_s$ without 
argument means $\alpha_s(\mu)$, and the number of colours and flavours equals $N_c=n_f=3$ in numerical expressions.

The Adler function can be written in the form 
\be
D(Q^2)=\frac{N_c}{12 \pi^2} \,\Big[1+\Delta_D(Q^2)\Big]\,.
\ee
The perturbative correction $\Delta_D^{\rm pert}(Q^2)$ is known to order $\alpha_s^4$ \cite{Baikov:2008jh,Baikov:2010je}, and is reprinted in appendix~\ref{app:series}.
With the above normalization, defining
\be
a_\mu\equiv\frac{\alpha_s(\mu)}{\pi}\,,
\ee
it reads in numerical form:
\begin{equation}
\Delta_D^{\rm pert}(Q^2) = a_Q + 1.64 a_Q^2 + 6.37 a_Q^3 + 49.08 a_Q^4+
\mathcal{O}(a_Q^5)\,.
\label{eq:DeltaDpert}
\end{equation}
The fourth-order term is as large as the third-order one 
at $Q\approx 1.20\,$GeV. To obtain this value we
use as input $\alpha_s(\mtau)=0.314$ in the three-flavour $\overline{\rm MS}$ scheme (corresponding to $\alpha_s(m_Z)=0.1180$ in the five-flavour scheme) and 
evolve it with five-loop accuracy to $Q$. The same strong coupling input will be used in all numerical results below. 

The above perturbative series diverges for any value of the coupling due to the rapid factorial growth of the coefficients. The leading asymptotic behaviour of the large-order coefficients arises from an  UV renormalon. This gives a sign-alternating divergent series which can be summed without uncertainties in principle. In practice, this turns out to be unnecessary, since its impact is known to be small numerically for the Adler function, which is the reason why the above series does not show 
sign alternation. The leading inevitable ambiguity arises from an IR renormalon ($u=2$ in the convention of ref.~\cite{Beneke:1998ui}, where more details of the present discussion can be found), giving a non-sign-alternating divergent behaviour and 
stemming from the IR region of loop integrals. The accuracy of a purely perturbative approach is therefore limited by 
\be
\label{eq:IRambiguity}
\delta \Delta_D^{\rm pert}(Q^2)=K C_{FF}(Q^2) \frac{\LMS^4}{Q^4} \,, 
\ee
where $K$ is an overall constant. 

In the broader, non-perturbative context of the OPE this intrinsic perturbative ambiguity is considered to be related to and cancelled against the uncertainty of the gluon condensate from its power-like (quartic) UV divergence \cite{Mueller:1984vh,Beneke:1993ee}. When the light-quark masses are set to zero, the OPE of the Adler function including power corrections to order $\LQCD^4/Q^4$ reads 
\be
D(Q^2)=\frac{N_c}{12 \pi^2} \left(C_1(Q^2)+\frac{C_{FF}(Q^2)}{Q^4} \langle \frac{\alpha_s}{\pi} F^2 \rangle +\mathcal{O}(\LQCD^6/Q^6) \right) . 
\label{eq:OPE}
\ee
At dimension-four shown here, there is only the vacuum matrix element of a single gauge-invariant operator involving the square of the gluon field strength tensor. It is convenient to employ the so-called scale-invariant gluon condensate 
\be
\langle \frac{\alpha_s}{\pi} F^2 \rangle \equiv 
\frac{\beta(\alpha_s)}{\pi \beta_0 \alpha_s} 
\langle 0| F^a_{\mu \nu} F^{a \mu \nu} | 0 \rangle\,,
\ee
which has no anomalous dimension. The short-distance coefficients $C_1(Q^2)=1+\Delta_D^{\rm pert}(Q), C_{FF}(Q^2)$ are given by perturbative series. $C_1(Q^2)$ follows from eq.~\eqref{eq:DeltaDpert}. $C_{FF}(Q^2)$ is known to order 
$\alpha_s^2$ \cite{Harlander1998,Bruser:2024zyg}, the explicit coefficients are summarized in appendix~\ref{app:series}. The OPE consistency requirement that the leading IR renormalon ambiguity \eqref{eq:IRambiguity} cancels with the gluon condensate fixes the coefficient of $\delta \Delta_D(Q^2)$ in terms of the gluon condensate short-distance coefficient and a single-number $K$, 
related to the ambiguous gluon condensate, which is $\mu$-independent due to the choice of the scale-invariant condensate, but depends on the scheme in such a way that $K \LMS^4$ is invariant. 

\subsection{Gradient-flowed gluon condensate}

In the standard picture above, the value of the gluon condensate is often assumed to depend on the prescription one envisages for summing the factorially divergent 
$\overline{\rm MS}$-scheme short-distance coefficients. Here we approach the problem from the other end. That is, we begin by providing a well-defined gluon condensate, which depends on an explicit cut-off that regulates the power 
divergences. In a second step, we replace the ambiguous gluon condensate by the gradient-flowed one, which induces a well-defined, IR-renormalon free short-distance coefficient. 

It is crucial to maintain gauge invariance of the regularization, in order 
to avoid artificial $1/Q^2$ power corrections. The regularization scheme should also be tractable in the sense that loop calculations can be performed systematically with reasonable effort. Finally, since non-perturbative calculations must eventually be done with lattice QCD, one must be able to take the continuum limit $a\to0$ of the lattice simulations at finite value of the cut-off that defines the condensates in order to combine the renormalized non-perturbative condensates with continuum perturbative calculations.

All three requirements are met by the gradient flow \cite{Luscher:2010iy,Luscher:2011bx}. 
In this framework, one defines the (Euclidean) 
flowed gluon field $B_{\mu}(t,x)=B_{\mu}^a(t,x) T^a$ by the flow equation 
\be
\partial_t B_{\mu}(t,x)=\mathcal{D}_{\nu} G_{\nu\mu}(t,x)+\alpha_0 \mathcal{D}_{\mu} \partial_{\nu} B_{\nu}(t,x)
\ee
with initial condition $B_{\mu}(t=0,x)=A_{\mu}(x)$, where $A_\mu(x)$ is the usual gluon field, $\mathcal{D}_{\mu}=\partial_{\mu}+g_0 [B_{\mu}, \cdot]$ and $\alpha_0$ is a gauge fixing parameter. Here $t$ is the flow ``time'', whose mass dimension is ${\rm dim} [t]=-2$. The field strength $G_{\mu \nu}(t, x)=G^{a}_{\mu \nu}(t,x)T^a$ of the flow field is 
\be
G_{\mu \nu}(t, x)=\partial_{\mu} B_{\nu}(x)-\partial_{\nu} B_{\mu}(x)+g_0 [B_{\mu}(x), B_{\nu}(x)] \,.
\ee
At zeroth order in the strong coupling and in Feynman gauge $\alpha_0=1$,  
the solution to the flow equation reads
\be B_\mu(t,x) = \int d^dy \,K(t, x-y)A_\mu(x)\,,
\qquad 
K(t, z) = \frac{e^{- z^2/(4 t)}}{(4\pi t)^{d/2}}\,,
\label{eq:treekernel}
\ee
which suggests that the gradient flow smears the singular, distribution-valued 
$A^\mu$-field with a Gaussian of variance $\sqrt{2 t}$, making gradient-flowed fields and local operators less singular than the usual ones. Indeed, we will find that 
$1/\sqrt{t}$ serves as a ``Wilsonian'' ultraviolet cut-off. A crucial feature of the gradient-flow regularization is that it preserves gauge-invariance, in contrast to explicit cut-offs of momentum integrals. 

The gradient-flowed gluon condensate is simply related to the ``action density'' 
\be
E(t)=\langle \frac{g_0^2}{4} G^a_{\mu \nu} G^a_{\mu \nu} \rangle(t)\,,
\ee
which is already being routinely calculated on the lattice for the purpose of scale setting \cite{FlavourLatticeAveragingGroupFLAG:2024oxs} and the determination of the strong coupling \cite{DallaBrida:2019wur,Hasenfratz:2023bok,Wong:2023jvr,Schierholz:2024lge,Larsen:2025wvg}. 
It was shown that $E(t)$ is UV finite and renormalization-scale invariant \cite{Luscher:2011bx}. At small flow times $1/t \gg {\LMS^2}$, the gradient-flowed operators admit themselves an OPE, resp. small flow-time expansion, in terms 
of short-distance coefficients in the $\overline{\rm MS}$ scheme and the same 
ambiguous condensates of the usual unsmeared fields as was the case for the Adler function:
\be
\frac{E(t)}{\pi^2}=\frac{Y_1(t)}{t^2}+Y_{FF}(t) \,\langle \frac{\alpha_s}{\pi} F^2\rangle+\mathcal{O}(t \LQCD^6)\,. 
\label{eq:smallflowtimeexpansion}
\ee
The $1/t^2$ term originates from the quartic power-UV divergence (recall that $t$ has mass dimension $-2$) and corresponds to mixing of the gluon operator $F^2$ with the unit operator in the OPE. The small flow-time expansion coefficients $Y_1(t)$, $Y_{FF}(t)$ are known at $\mathcal{O}(\alpha_s^3)$ \cite{Harlander:2016vzb} and $\mathcal{O}(\alpha_s^2)$ \cite{Harlander:2020duo}, respectively. Their explicit expressions are given in appendix~\ref{app:series}. 

\subsection{Renormalon subtraction and gradient-flowed OPE}

Universality of the gluon condensate and consistency of the small-flow time expansion require that $Y_1(t)$ exhibits an IR renormalon factorial divergence which is tied to the same ambiguous standard gluon condensate as the one in 
$C_1(Q^2)$. We therefore replace the standard condensate $\langle \frac{\alpha_s}{\pi} F^2 \rangle$ in the Adler function OPE \eqref{eq:OPE} by the unambiguous gluon condensate $E(t)$. To this end, we solve eq.~\eqref{eq:smallflowtimeexpansion} for 
\be
\langle \frac{\alpha_s}{\pi} F^2\rangle=\frac{1}{Y_{FF}(t)} \left(\frac{E(t)}{\pi^2}-\frac{Y_1(t)}{t^2} \right)+\mathcal{O}(t \LQCD^6) \label{eq:GCdim4}
\ee
to obtain the new OPE for the Adler function:
\begin{eqnarray}
D(Q^2)&=&\frac{N_c}{12\pi^2} \,\bigg[ \left(C_1(Q^2)-\frac{1}{t^2 Q^4} r(Q^2, t) Y_1(t) \right)
\nonumber\\
&&+\,\frac{r(Q^2, t)}{Q^4}  \frac{E(t)}{\pi^2}+\mathcal{O}(t \LQCD^6/Q^4, \LQCD^6/Q^6) \,\bigg] , 
\label{eq:newOPE}
\end{eqnarray}
introducing the ratio of the gluon condensate short-distance coefficients
\be
r(Q^2, t)= \frac{C_{FF}(Q^2)}{Y_{FF}(t)} \,. 
\label{eq:r}
\ee
We refer to eq.~\eqref{eq:newOPE} and more generally an OPE where all local operator matrix elements have been replaced by their gradient-flowed counterparts as the {\em gradient-flowed OPE}. 

Eq.~\eqref{eq:newOPE} is the central formula of this approach. We discuss the terms in square brackets separately. 
\begin{itemize}
\item[(1)] The two terms in round brackets in the first line of the equation represent the original Adler function series and a subtraction term. Both together are referred to as the subtracted Adler function series. The leading IR renormalon divergence of the original Adler function is eliminated in this combination. This can be seen as follows. The short-distance coefficient $Y_1(t)$ in the small flow-time expansion~\eqref{eq:smallflowtimeexpansion} has the leading ($u=2$) renormalon uncertainty 
\be
\delta Y_1(t)=K Y_{FF}(t) \LMS^4 t^2\,, 
\label{eq:Y1u2}
\ee
which parallels eq.~\eqref{eq:IRambiguity} for the Adler function with the same constant $K$ due to the universality of the gluon condensate 
$\langle \frac{\alpha_s}{\pi} F^2 \rangle$. Due to   the $1/(t^2 Q^4)$ prefactor of the subtraction term, the leading infrared large-order behaviour cancel, and 
one therefore expects the subtracted Adler function series to exhibit  smaller perturbative corrections. 
\item[(2)] The first term in the second line of eq.~\eqref{eq:newOPE} is the 
gradient-flowed gluon condensate contribution. Here $E(t)$ is an unambiguously defined non-perturbative quantity, given by the continuum limit of the action density computed at flow time $t$ in lattice QCD. 
\end{itemize}

The gradient-flowed OPE therefore solves both problems of the original OPE by replacing divergent perturbative series and ill-defined condensates with a convergent series (as far as the leading IR renormalon is concerned) and a non-perturbatively defined condensate. In eq.~\eqref{eq:newOPE}, $1/\sqrt{t}$ can be regarded alternatively as a gauge-invariant IR cutoff on perturbative loop momentum integrals, as the subtraction term (the second term inside the parentheses) vanishes 
in the formal limit $1/\sqrt{t} \to 0$, or a UV cut-off on the local operators. 
The cut-off, resp. gradient-flow time $t$ in eq.~\eqref{eq:newOPE}   should be chosen to satisfy\footnote{The factor of 1/2 is motivated by eq.~\eqref{eq:treekernel}.}
\begin{equation}
{\LMS^2} \ll 1/(2t) \ll Q^2\,.
\label{eq:twindow}
\end{equation}
The condition ${\LMS^2} \ll 1/(2t)$ is necessary to ensure the validity of the small flow time expansion. As a consequence the condensate $E(t)$ is not simply of order $\LMS^4$, but of order $1/t^2$, which is parametrically larger. However, these 
cut-off terms cancel between the subtraction terms and the gluon-condensate 
terms. In order to maintain an OPE in $1/Q^2$, the second condition $1/(2t) \ll Q^2$ 
is required.  

Let us note that $C_{FF}$ and $Y_{FF}$ and therefore $r$ contain themselves a leading IR renormalon singularity at $u=1$ related to dimension-six operators in their OPE, resp. small flow-time expansion. We discuss the effect of dimension-six operators and the role of the renormalons in $r$ and the truncation of $r$ in sec.~\ref{sec:dim6} and appendix~\ref{app:truncationorder}, respectively.

\subsection{Illustration in the large-$\beta_0$ approximation}

We exemplify how the renormalon cancellation in the gradient-flowed OPE of the Adler function is realized in the so-called large-$\beta_0$ approximation. Technically, this approximation amounts to calculating the leading terms in the $n_f\to -\infty$ limit, consisting of fermion bubble insertions into a single gluon propagator 
of the two-loop diagrams for the current correlation function, 
and then substituting $-\frac{2 n_f}{3} \to -4\pi\beta_0$  \cite{Beneke:1994qe,Neubert:1994vb,Ball:1995ni}. 

We define the Borel transform of 
\be
C_1(Q^2)=1+\sum_{n=0}^{\infty} d_n \alpha_s^{n+1}\,,
\ee
as 
\be
B_D(u)=\sum_{n=0}^{\infty} \frac{d_n}{n!} \lp \frac{u}{-\beta_0} \rp^{\!n} \,.
\ee
With this convention, the IR renormalon singularities of the Adler function series are located at integer $u=2,3,\ldots$. In the large-$\beta_0$ approximation for the coefficients $d_n$, the Borel transform of the Adler 
function was computed in refs.~\cite{Beneke:1992ch,Broadhurst:1992si}. It 
can be expressed in the form \cite{Broadhurst:1992si} 
\be
B_D(u)=\frac{32}{3 \pi} \lp \frac{\mu^2 e^{5/3}}{Q^2} \rp^{\!\!u}  \frac{1}{2-u} \sum_{k=2}^{\infty} \frac{(-1)^k k}{(k^2-(1-u)^2)^2}\,,
\label{eq:BorelD}
\ee
which makes the renormalon poles manifest.\footnote{The sum can be written in terms of the trigamma function \cite{Mikhailov:2023lqe,Laenen:2023hzu}. Explicit 
values of the first few $d_n$ can be inferred from Table~1 of ref.~\cite{Ball:1995ni} after adjusting the convention there to the present one.} Likewise, in the large-$\beta_0$ approximation the Borel transform 
of the $1/t^2$ term in the small-flow time expansion \eqref{eq:smallflowtimeexpansion} of the action density reads  
\cite{Suzuki:2018vfs} 
\be
B_E(u)=\sum_{n=0}^{\infty} \frac{e_n}{n!} \lp \frac{u}{-\beta_0} \rp^n 
=\frac{3}{4 \pi^3} \lp 2 t \mu^2 e^{5/3}\rp^u \Gamma(2-u)
\label{eq:BorelE}
\ee
with series coefficients $e_n$ from the expansion
\be
Y_1(t)|_{\text{large-$\beta_0$}}=\sum_{n=0}^{\infty} e_n \alpha_s^{n+1}\,.
\ee
Contrary to the Adler function, the action density series exhibits no 
UV renormalon poles at negative integer $u$. The IR renormalons at $u=2,3,\ldots$ are due to the Gamma function.

The large-order behaviours of the perturbative coefficients
generated by the $u=2$ IR renormalon are then given by
\be
d_n \approx \frac{e^{10/3}}{\pi} \lp\frac{\mu^2}{Q^2} \rp^{\!2} \left(-\frac{\beta_0}{2} \right)^{\!n} \,n!\,,
\label{dnasym}
\ee
\be
e_n \approx \frac{3 e^{10/3}}{8 \pi^3}  \lp 2 t \mu^2 \rp^{2} \left(-\frac{\beta_0}{2} \right)^{\!n} \,n!\,.
\label{enasym}
\ee
In the large-$\beta_0$ approximation, the Wilson coefficients $C_{FF}(Q^2)$
and $Y_{FF}(t)$ take their leading-order values
\be
C_{FF}(Q^2)=\frac{2 \pi^2}{3}, \qquad{} Y_{FF}(t)=1\,.
\ee
Therefore, in the subtracted perturbative series
\be
C_1(Q^2)-\frac{1}{t^2 Q^4}\frac{C_{FF}(Q^2)}{Y_{FF}(t)} Y_1(t)
=1+\sum_{n=0}^{\infty} \lp d_n-\frac{1}{t^2 Q^4} \frac{2 \pi^2}{3} e_n \rp \alpha_s^{n+1}(\mu^2)\,,
\ee
these divergent behaviours are indeed cancelled. Alternatively, using 
$\sum_{k=2}^\infty \frac{(-1)^k k}{(k^2-1)^2}=\frac{3}{16} $, it can be seen directly from the Borel transforms that 
\be 
B_D(u) - \frac{2 \pi^2}{3} \frac{1}{t^2 Q^4} B_E(u)
\ee
is non-singular at $u=2$.

We make some remarks. 
First, for the order-by-order cancellation in perturbation theory
the same renormalization scale $\mu$ should be used for the series expansion of $C_1(Q^2)$ and $Y_1(t)$.
Second, the factors $(\mu^2/Q^2)^2$ and $(t \mu^2)^2$ in~\eqref{dnasym}
and \eqref{enasym} arise, because logarithmic terms are effectively 
exponentiated at large orders---the $n$th order perturbative coefficients 
$d_n$ and $e_n$ are originally given by $n$th-order polynomials of\footnote{The factor 8 in $8t$ is by convention.} 
\be
L \equiv \log(Q^2/\mu^2), \quad{}
L_t \equiv \log(1/(8 t \mu^2))\,,
\label{LandLt}
\ee
respectively. 
Third, choosing a too large hierarchy $8 t Q^2 \gg 1$
is not preferable from the viewpoint of the renormalon cancellation.
While the cancellation formally holds for arbitrary $\mu$, $t$, and $Q$, 
it takes effect only at the perturbative order $n$ which satisfies $|(-2 L_t)^n/n!| \ll 1$, so that the approximation
\be
\sum_{k=0}^n  \frac{(-2)^k}{k!} L_t^k \approx  (8 t \mu^2)^2 
\ee
is valid.
Since the typical scale of the observable is $Q$, 
we assume $\mu \sim Q$ and thus $L_t \sim \log(1/(8 t Q^2))$. 
Therefore, a too large $8 t Q^2$ makes the renormalon cancellation 
happen only at very large order. This issue should be taken into account 
when choosing a value for $t$.

\section{Adler function}
\label{sec:adler}

We examine how the gradient-flowed OPE solves the IR renormalon problem and improves the theoretical accuracy of the (flavour non-singlet, $n_f=3$) Adler function, given the actual perturbative expansion coefficients.  As a reference energy value we adopt $Q=m_{\tau}=1.777$~GeV, the mass of the $\tau$ lepton. 
Correspondingly, the strong coupling in the three-flavour scheme at scale $m_\tau$, $\alpha_s(m_{\tau})=0.314$, is used, which is derived from $\alpha_s(M_Z)=0.1180$ by five-loop evolution and decoupling of the bottom and charm quarks. 

We evaluate the short-distance coefficients $C_1, C_{FF}, Y_1, Y_{FF}$ in perturbation theory. The relevant perturbative series are 
currently available to the following orders: N$^4$LO ($\mathcal{O}(\alpha_s^4)$) for $C_1(Q^2)$ \cite{Baikov:2008jh}, N$^2$LO
($\mathcal{O}(\alpha_s^3)$) for $Y_1(t)$  \cite{Harlander:2016vzb,Artz:2019bpr}, and N$^2$LO ($\mathcal{O}(\alpha_s^2)$) for $C_{FF}(Q^2)$ \cite{Harlander1998} and $Y_{FF}(t)$  \cite{Harlander:2020duo}. 
In case of $C_{FF}(Q)$ and $Y_{FF}(t)$, the highest available order is next-to-next-to-leading order (NNLO) and consequently we evaluate the ratio $r$~\eqref{eq:r} to $\mathcal{O}(\alpha_s^2)$. The subtracted Adler function coefficient at order $\alpha_s^n$ is then obtained by expanding $C_1 - r Y_1/(t^2 Q^4)$ in eq.~\eqref{eq:newOPE} to the $n$th order. The action density $E(t)$ is constructed non-perturbatively from lattice data as discussed below.

In order to illustrate the expected leading IR renormalon cancellation in higher orders, in addition to the exactly known perturbative coefficients of $C_1$, $Y_1$, we use a model for the Borel transform of the entire $C_1$, $Y_1$ series, developed in ref.~\cite{Beneke:2008ad} and generalized here to include $Y_1$. The model is based on an ansatz for the Borel transform that includes the leading UV and IR renormalon singularities and a low-order polynomial, which is matched to the exactly known perturbative coefficients. A detailed description is given in appendix~\ref{app:renormalonmodel}. The model allows us to generate all-order coefficients for $C_1$, $Y_1$, which match smoothly to the exactly known ones and incorporate the expected sign-alternating and fixed-sign asymptotic behaviours  in large orders. We emphasize that this model is employed only for illustration. The main results and improvements due to gradient-flow subtraction rely only on the exactly known coefficients. Results making use of the model will be presented in fainter  colour in figures. 

\subsection{Renormalon subtraction}

We first focus on the first term of eq.~\eqref{eq:newOPE}, 
\be
C_1(Q^2)-\frac{1}{t^2 Q^4} r(Q^2, t) Y_1(t) \equiv 1+\Delta_D^{\rm pert}(Q^2,t)\,,
\ee
which we refer to as the (gradient-flow) subtracted perturbative Adler function. We also define the series expansion in 
$a_Q=\alpha_s(Q)/\pi$ summed to order $N$ by
\be
\Delta_D^{\rm pert}(Q^2,t) = \sum_{n=1}^N c_n(t) a_Q^n\,.
\label{eq:partialDpert}
\ee
The flow time, which defines the subtraction scale, should satisfy the condition ${\LMS^2} \ll 1/(2t) \ll Q^2$ as discussed above, and we adopt $t=20/(8 Q^2)$ as the default value. To display the dependence on the subtraction scale, we also consider $t=15/(8 Q^2)$ and $t=30/(8 Q^2)$. With these choices, and since the renormalization scale is set to $\mu=Q$, the coefficients $c_n(t)$ are $\mu$ and $Q$ independent. 

\begin{table}[t]
\centering
\begin{tabular}{r|r| r r r}
order in $\alpha_s$  & Unsubtracted & $t=20/(8 Q^2)$ & $t=15/(8 Q^2)$ & $t=30/(8 Q^2)$ \\
\hline
1 & 1.00000 & 0.920000 & 0.857778 & 0.964444 \\
2 & 1.63982 & 0.935656 & 0.48003 & 1.29442 \\
3 & 6.37101 & 0.686040 & -1.95467 & 3.16625 \\ \cdashline{3-5}
4 & 49.0757 & 4.62963 & -6.79875 & 19.4893 \\  \cdashline{2-2}
5 & 283.000 & -90.6174 & -143.903 & 0.914328 \\
6 & 3279 & -112 & -370 & 464 \\
7 & 1.84$\times10^4$ & -1.43$\times10^4$ & -1.53$\times10^4$ & -1.09$\times10^4$ \\
8 & 3.88$\times10^5$ & 4.96$\times10^4$ & 4.97$\times10^4$ & 6.67$\times10^4$ \\
9 & 8.39$\times10^5$ & -2.92$\times10^6$ & -2.85$\times10^6$ & -2.88$\times10^6$ \\
10 & 8.40$\times10^7$ & 3.89$\times10^7$ & 4.00$\times10^7$ & 3.83$\times10^7$ \\
11 & -5.47$\times10^8$ & -1.13$\times10^9$ & -1.12$\times10^9$ & -1.15$\times10^9$ \\
12 & 3.42$\times10^{10}$ & 2.59$\times10^{10}$ & 2.61$\times10^{10}$ & 2.57$\times10^{10}$ \\
\end{tabular}
\caption{Perturbative coefficients $c_n(t)$ for the Adler function (unsubtracted) and the gradient-flowed subtracted Adler function with different choices of the flow time ($8 Q^2 t=20, 15, 30$). 
Numbers above the dashed line refer to exactly known coefficients, those below are estimated based on the Borel-transform model discussed at the beginning of sec.~\ref{sec:adler} and in appendix~\ref{app:renormalonmodel}. 
}
\label{tab:1}
\end{table}

\begin{figure}[t] 
\begin{center}
\includegraphics[width=12cm]{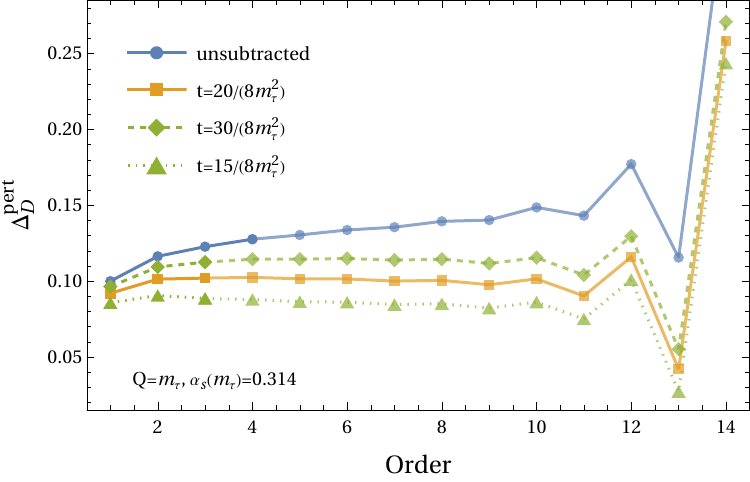}
\end{center}
\caption{
The perturbative part $\Delta_D^{\rm pert}(Q^2,t)$ 
of the Adler function at $Q=m_{\tau}$, 
$\alpha_s(m_\tau)=0.314$
summed to perturbative order $N$. Shown are the 
unsubtracted and gradient-flow subtracted series 
(flow times $8 m_{\tau}^2 t=20, 30, 15$). 
The fainter colours indicate that these results 
rely on the estimated perturbative coefficients, i.e. 
beyond the 4th order for the ``unsubtracted'' series 
and beyond the 3rd
order for the subtracted series.}
\label{fig:fixedorderadler}
\end{figure}

In table~\ref{tab:1}, we list the coefficients $c_n(t)$
of the perturbative part of the Adler function and compare 
the standard, unsubtracted perturbative series to the 
gradient-flow subtracted one for the above three values 
of $t$. We notice that the growth of perturbative 
coefficients in low and intermediate orders (say, $1-7$) is alleviated in the subtracted perturbative series, which we attribute to the successful subtraction of the 
leading $u=2$ IR renormalon. Since the nominally leading 
asymptotic behaviour is due to the UV renormalon at $u=-1$, 
which is not affected by the subtraction, at 
higher order (say, 8-th order) the sign-alternating factorial divergence becomes visible in all cases, see also fig.~\ref{fig:fixedorderadler}. 

The effect of renormalon subtraction is better illustrated by plotting the dependence of the partial sums \eqref{eq:partialDpert} on the truncation order $N$, as shown in 
fig.~\ref{fig:fixedorderadler}. The standard Adler function exhibits a continuous growth in intermediate orders. 
It is remarkable that in contrast 
the subtracted Adler function with the choice of $t=20/(8 m_{\tau}^2)$ hardly changes beyond the second order 
in perturbation theory. Both transit to the sign-alternating behaviour at high orders. This behaviour is stable with respect to different choices of the flow time $t$ around
$t=20/(8 Q^2)$. 

This demonstrates that gradient-flow subtraction works exceptionally well for the leading IR renormalon divergence, resulting in a well-behaved Adler function series at the relevant intermediate orders in perturbation theory. The subtraction moves a ($t$-dependent) part of the standard series into the gradient-flowed gluon condensate/action density. In the following subsection, we examine this non-perturbative contribution and investigate how the $t$-dependence seen in fig.~\ref{fig:fixedorderadler} cancels after combining the above perturbative series with the gluon condensate.

\subsection{Gradient-flowed gluon condensate}
\label{sec:lattice}

We therefore consider the second term of eq.~\eqref{eq:newOPE}, 
which requires the non-perturbative evaluation of the action density $E(t)$ on the lattice. This term has the following significance. First, it includes the non-perturbative
contribution of $\mathcal{O}(\LQCD^4)$, as can be seen from
eq.~\eqref{eq:smallflowtimeexpansion}.
As a result, it unambiguously provides the $\mathcal{O}(\LQCD^4)$ correction to the Adler function.
This correction is usually difficult to determine due to the ($u=2$) renormalon divergence.
Even in the case the renormalon ambiguity is fixed by the principal-value prescription of the Borel transform, 
the extraction of the gluon condensate seems to require  
perturbative calculations to very large orders  \cite{Bali:2014sja,Ayala:2020pxq}.
The gradient-flow formulation avoids these issues and allows us to reach non-perturbative precision with comparatively little effort.
Second, this term is responsible for cancelling 
the $t$, resp. explicit factorization scale dependence, introduced 
by the first term of eq.~\eqref{eq:newOPE} (up to higher order $t$ dependence corresponding to higher-dimensional operators).
This cancellation happens since the Adler function
is of course independent of the flow time $t$.

To proceed, non-perturbative data of $E(t)$ is required. Unfortunately, there seem to exist no public lattice data of the continuum limit of $E(t)$ that entirely covers the range we are interested in, say $\sqrt{t} \in [0.1, 0.4]$~fm (considering $Q$ between 1 and 3~GeV).  In the present study, we instead employ existing lattice data to approximate the $t$-dependence as follows. There are two well-known scales called $t_0$ and $w_0^2$, defined through $E(t)$:
\be
t^2 E(t)|_{t=t_0}=0.3, \quad{} t \frac{d}{dt} \lp t^2 E(t) \rp \bigg|_{t=w_0^2}=0.3 .
\ee
We consider the linear-order approximation of $E(t)$ around $t=t_0$, 
consistent with these conditions:
\be
t^2 E(t)|_{\rm lin}=0.3+\frac{0.3}{w_0^2} (t-t_0) \label{linearmodel} .
\ee
Using the FLAG averages \cite{FlavourLatticeAveragingGroupFLAG:2024oxs} for these parameters (with the errors neglected),
\be
\sqrt{t_0}=0.14292~{\rm fm}, \quad{} w_0=0.17256~{\rm fm},
\ee
we can obtain $E(t)$ at an arbitrary $t$ based on eq.~\eqref{linearmodel}.

\begin{figure}[t] \centering
\begin{center}
\hspace*{0.4cm}\includegraphics[width=11cm]{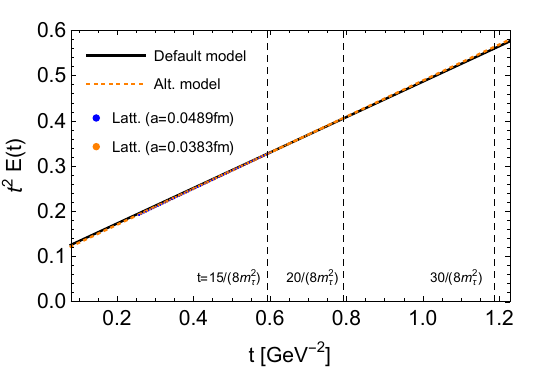}\\[0.4cm]
\hspace*{-0.12cm}\includegraphics[width=10.55cm]{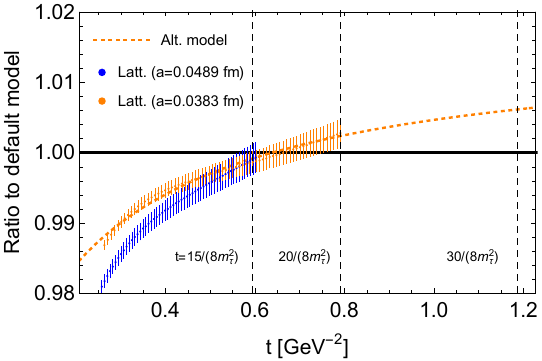}
\end{center}
\caption{Upper panel: Non-perturbative action density. ``Default model'' and ``Alt. model'' refer to eq.~\eqref{linearmodel} and eq.~\eqref{eq:alternativemodel}, respectively. For the lattice data, 
see text.
The reference scales associated with the action density
are given in units of GeV${}^{-2}$ by
$t_0 =0.525~\mbox{GeV}^{-2}$ and $w_0^2=0.765~\mbox{GeV}^{-2}$. 
Lower panel: Ratios of the lattice data to the linear model eq.~\eqref{linearmodel}.
}
\label{fig:NPactiondensity}
\end{figure}

In fig.~\ref{fig:NPactiondensity} 
we compare this parameterization with actual lattice data at finite lattice spacings from refs.~\cite{Mohler:2017wnb,Bruno:2014jqa} (reweighted according to ref.~\cite{Kuberski:2023zky}).\footnote{We thank Alejandro Saez Gonzalvo and Alberto Ramos for providing us with these lattice data.} 
Good agreement is confirmed in the wide range 
$0.2~{\rm GeV}^{-2} \lessim \,\,t \lesssim 0.8~{\rm GeV}^{-2}$
within a few per cent deviations. We also employ another model obtained from a linear fit (minimizing $\chi^2$ with the statistical errors taken into account) to the finer lattice data ($a=0.0383$~fm) shown in the figure. This gives 
\be
t^2 E(t)|_{\text{fine latt.}}=
0.29929+0.39855 \,\mbox{GeV}^2\,(t-t_0)\,,
\label{eq:alternativemodel}
\ee
and is displayed as the dashed orange line in the 
figure. 

In the study below, the linear model \eqref{linearmodel} is employed by default. 
The difference to the alternative model \eqref{eq:alternativemodel} is 
used to access the systematic uncertainty of the approximations, including the extrapolation outside the range where the lattice data are available.

\subsection{Non-perturbative Adler function in gradient-flowed OPE}
\label{sec:adlernonpertresults}

With both terms of eq.~\eqref{eq:newOPE} in place, we next analyze the non-perturbative Adler function, which consistently includes the dimension-4 gluon condensate on top of the truncated perturbative expansion of the unit operator.

\subsubsection{$t$ independence}

\begin{figure}[t] \centering
\begin{center}
\includegraphics[width=14cm]{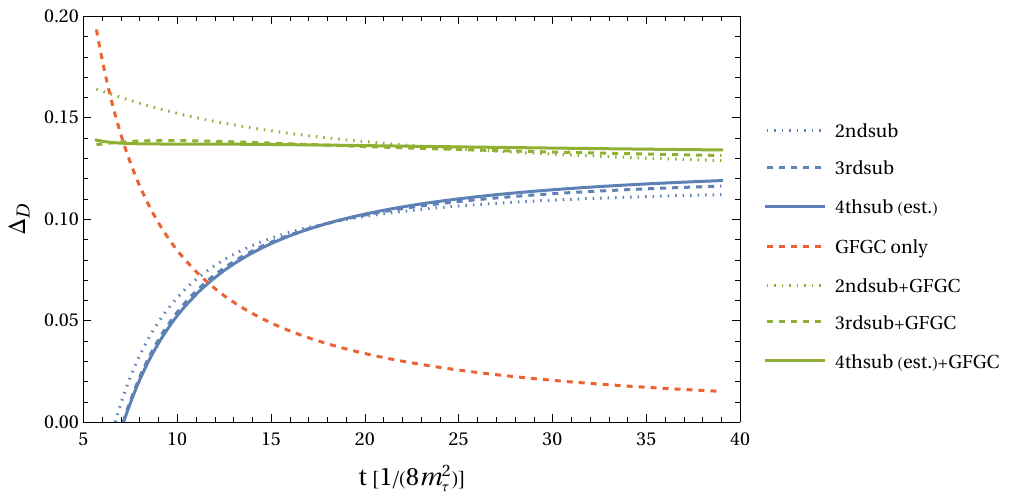}\\[0.2cm]
\includegraphics[width=14cm]{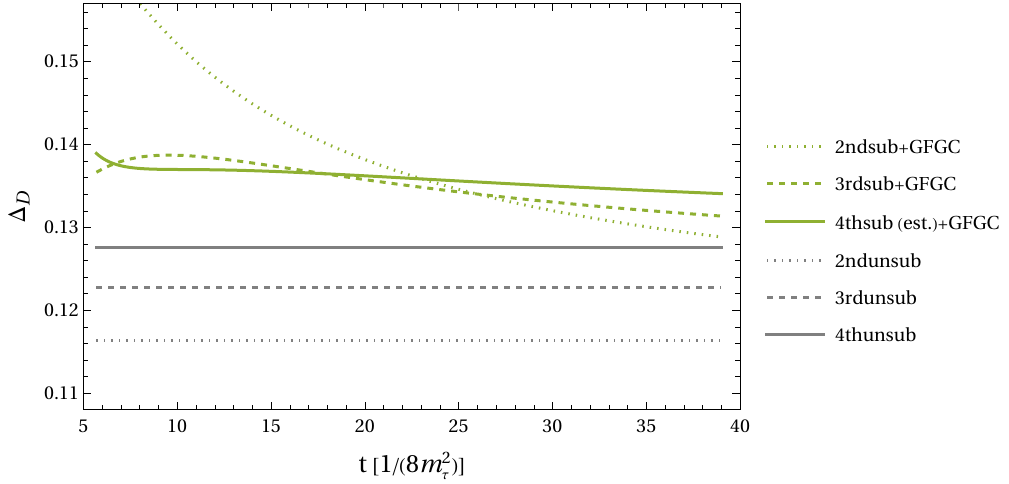}
\end{center}
\caption{
$t$-dependence of the Adler function at $Q=m_{\tau}$, $\alpha_s(m_\tau)=0.314$. Shown are the perturbative subtracted contribution (blue), the gradient-flowed gluon-condensate contribution (GFGC, red-dashed) and the sum (green) 
at the second, third and fourth order in perturbation theory. The 
lower panel shows a zoom-in that compares the non-perturbative to the standard (unsubtracted) perturbative approximation at second, third, and fourth order for the perturbative part.
}
\label{fig:tdep}
\end{figure}

We first examine the cancellation of $t$ dependence between the
subtracted unit-operator term and the gradient-flowed (GF) gluon-condensate term
in eq.~\eqref{eq:newOPE}. In the upper fig.~\ref{fig:tdep}, 
we show the $t$ dependence of both terms separately, in blue and red, 
respectively. For the subtracted unit operator, we indicate the perturbative truncation order by dotted (2nd), dashed (3rd) and 
solid (4th).\footnote{While the 4th order expansion coefficient of the Adler function $C_1$ in standard perturbation theory is known exactly, the subtraction term $Y_1$ is known exactly only to 3rd order. In this case the estimate for $e_3$ from the Borel function model is used as indicated by the qualifier ``(est.)'' in the figure legend. }  The sum of the two terms gives the non-perturbative Adler function and is shown by the green lines.

While the subtracted term and the GF gluon-condensate term separately exhibit strong $t$ dependence, 
it is largely cancelled in the sum. Given that over the displayed range of $t$, the two terms separately vary by a factor of several, this is a highly non-trivial consistency check, since it implies that the non-perturbative lattice data is compatible with its small flow-time expansion to a very high degree.

For too small $t$, the renormalon cancellation 
works only at rather large perturbative truncation order, as mentioned earlier. Effectively, 
the UV cut-off on the contributions included in the condensate is large compared to the strong interaction scale. The condensate 
increases with the fourth power of this cut-off and becomes large, but is effectively including perturbative contributions. 
On the other hand, too large $t$ is not preferable from the 
viewpoint of the validity and convergence of the small flow-time expansion. Taking into account these considerations and the observed $t$ dependence, we focus on the range $15 \leq 8 m_{\tau}^2 t \leq 30$.

The bottom panel of fig.~\ref{fig:tdep} magnifies the behaviour of the non-perturbative Adler function in the GF OPE (green lines) and compares it to the  standard perturbative series for
$C_1(Q^2)$, which of course does not depend on $t$.
The non-perturbative result is seen to be more stable than
the standard result against higher-order perturbative corrections
in the $t$ range mentioned above.

\subsubsection{Uncertainty estimates}

Motivated by this observation of successful renormalon subtraction and its stability against $t$ variation and higher-order corrections,
we compare the theoretical error of the non-perturbative Adler function to the standard perturbative truncation. 

The uncertainties are estimated as follows. First, for the perturbative uncertainty we vary the renormalization scale around $\mu=Q$ 
by a factor of $\sqrt{2}$, i.e. between $\mu=\sqrt{2} Q$ and 
$\mu=Q/\sqrt{2}$.  Second, the result should ideally be independent 
of the chosen $t$, but can depend on $t$ due to missing higher 
orders in perturbation theory and the OPE. We adopt the 
$t$-values $t=15/(8 Q^2)$ and $30/(8 Q^2)$ to estimate the uncertainty. Third, we estimate the uncertainty arising from 
modelling the non-perturbative data of $E(t)$ and its 
extrapolation outside the $t$-range for which it is available by taking plus/minus 
the difference between the two parameterizations discussed in sec.~\ref{sec:lattice} at $t=20/(8 Q^2)$.
Central values are given with $\mu=Q$, $t=20/(8 Q^2)$ and 
the linear model  \eqref{linearmodel} for $E(t)$.
We vary each parameter independently to examine
the variation around the central values.
Finally all the uncertainties are combined in quadrature.

\begin{table}[t]
\centering
\begin{tabular}{r|c|l|}
Order & Standard PT & Gradient-flowed OPE [to dim-4]\\
\hline
&&\\[-0.4cm]
NLO & $0.100^{+0.025}_{-0.016}$ & \;\;$0.135^{+0.023}_{-0.015} (\mu)^{+0.013}_{-0.012} (t)$\\
NNLO & $0.116^{+0.010}_{-0.010}$ & \;\;$0.138^{+0.004}_{-0.006} (\mu)^{0.005}_{-0.006} (t)$\\
N$^3$LO & $0.123^{+0.006}_{-0.006}$& \;\;$0.1358^{+0.0000}_{-0.0018} (\mu)^{+0.0017}_{-0.0027} (t)$\\
N$^4$LO & $0.128^{+0.006}_{-0.005}$ & \;\;$0.1362^{+0.0000}_{-0.0007} (\mu)^{+0.0005}_{-0.0012} (t)$ (est.) \\
\end{tabular}
\vspace*{0.2cm}
\caption{Numerical value of $\Delta_D(m_\tau^2)$ ($\alpha_s(m_\tau)=0.314$) for the first four perturbative orders. The uncertainty of ``standard perturbation theory'' is computed from varying the renormalization scale $\mu$ as described in the text. The result from the gradient-flowed OPE includes the dimension-4 gluon-condensate contribution. The first error is due to scale variation, the second from the variation of the small flow-time subtraction scale. In case of the gradient-flowed OPE  the N$^4$LO result includes an estimate (est.) of $e_3$ from the Borel function model.}
\label{tab:2}
\end{table}

The result of this analysis is summarized in tab.~\ref{tab:2} 
for perturbative truncation orders NLO to fourth order (N$^4$LO) 
for the unit operator. The third uncertainty from non-perturbative data is negligible at $Q=m_\tau$ (at most 0.0001) and therefore omitted. We observe a progressive reduction of the dependence on $\mu$ and $t$ as the perturbative order is raised. 
Adding uncertainties we obtain 
\begin{align}
\Delta_D(m_{\tau}^2)
&=\begin{cases}
 0.135^{+0.026}_{-0.019} & (\text{LO}) \\
 0.138^{+0.007}_{-0.009}  & (\text{NLO}) \\
0.1358^{+0.0017}_{-0.0033} & (\text{N$^3$LO}) \\
0.13621^{+0.0006}_{-0.0014} & (\text{N$^4$LO, est.})
\end{cases} \label{AdlerMtauGFOPE}
\end{align}
for $\alpha_s(m_{\tau})=0.314$. At third and fourth order the theoretical uncertainty is reduced by a factor of five relative to the previous approximations based on standard perturbation theory (in the latter case, the uncertainty is computed from scale variation alone). Since the gradient-flow OPE result is non-perturbative, one should also consider the possible effect of higher-dimension condensates in the OPE, which are poorly known. While we cannot exclude that sizeable values of the dimension-six condensates shift the above numbers, the power-like residual $t$-dependence incurred by the truncation of the OPE to dimension four is included in the $t$-variation estimate.

\begin{figure}[t] \centering
\begin{center}
\hspace*{-0.5cm}\includegraphics[width=11cm]{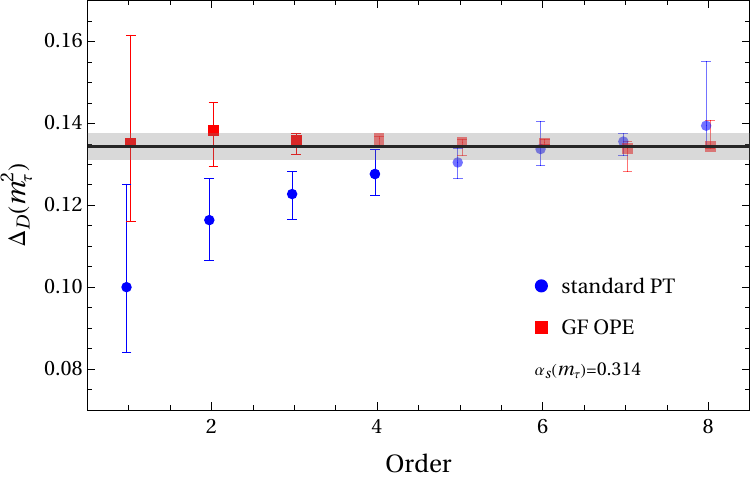}
\end{center}
\caption{Dependence of the Adler function at $Q^2=m_{\tau}^2$ in standard perturbation theory (blue) and the gradient-flowed OPE (red) on the truncation order of the unit operator short-distance coefficients. The fainter points rely on the Borel function model to extrapolate the series expansions to higher order than exactly known. The black horizontal line/grey band display the Borel sum/ambiguity of the standard perturbative series.}
\label{fig:fixedorderDmtau}
\end{figure}

In fig.~\ref{fig:fixedorderDmtau}, we compare the non-perturbative Adler function in the gradient-flowed OPE (red squares) to standard perturbation theory (blue points) as the truncation order of the perturbative part is increased. Error bars are computed as described above. 
The extrapolation to higher orders than known exactly uses the Borel transform model for the unsubtracted Adler function series. The black line represents the Borel integral of this Borel transform and the grey band the estimate of the intrinsic uncertainty of a purely perturbative approach based on the ambiguity of Borel summation (computed from the imaginary part of the Borel integral divided by $\pi$). The figure highlights two important observations: 
\begin{itemize}
\item[(1)] The value of the gradient-flowed OPE is very stable starting already at very low orders. Higher-order computations reduce the theoretical uncertainty but small uncertainties are already obtained at the third order. This should be contrasted to standard perturbation theory. 
At the highest orders shown the uncertainty increases again. This is due to the onset of sign-alternating UV renormalon divergence. This could be dealt with by conformal mapping techniques, but is not relevant in practice, since the fourth or fifth order approximation is sufficient. 
\item[(2)]  The non-perturbative GF OPE reaches precision below the intrinsic ambiguity of the standard perturbative treatment from its factorial divergence, which proves that one can go beyond it, that is, to non-perturbative accuracy with this method. At the same time, the fact that (in the adopted Borel function model) standard perturbation theory and the GF OPE converge within errors within the grey band at the  sixth and seventh order suggests that the ``true" gluon condensate (whatever this means) is not larger than the estimated perturbative ambiguity. For reference, the so-called SVZ canonical value $\langle \frac{\alpha_s}{\pi} F^2\rangle=0.012\,\mbox{GeV}^{4}$ \cite{Shifman:1978bx} of the gluon condensate employed in many phenomenological studies amounts to a contribution  0.0079 to $\Delta_D(m_\tau^2)$ 
to be compared to the width 0.006 of the grey band. 
\end{itemize}

\subsubsection{Energy dependence}

\begin{figure}[t] \centering
\begin{center}
\includegraphics[width=11cm]{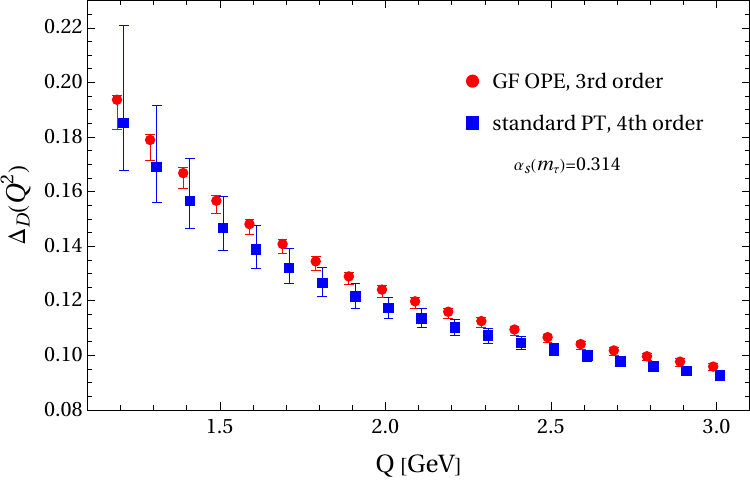}
\end{center}
\caption{
The Adler function $\Delta_D(Q^2)$ as a function of $Q$. 
The gradient-flow OPE result (red) with the NNNLO perturbative series and $E(t)$ compared to the standard perturbative results at N$^4$LO (blue) are shown.  $\alpha_s(m_{\tau})=0.314$ is used. The error bars are computed as for the previous figure and described in the text.
}
\label{fig:AdlerQsqdep}
\end{figure}

So far we fixed $Q=m_{\tau}$. We now study the Adler function at different energies $Q$, in a manner parallel to the above analysis.
Fig.~\ref{fig:AdlerQsqdep} gives the results obtained in standard perturbation theory and the gradient-flowed OPE at the highest order at which the perturbative coefficients are presently known exactly. The error bars are computed as described above and include the error estimate from the parameterization of the non-perturbative action density. As $Q$ becomes smaller, the subtraction flow-time $t=20/(8 Q^2)$ increases and moves into the regime of fig.~\ref{fig:NPactiondensity}, where lattice data is extrapolated and the difference between the two parameterizations increases. Nevertheless, only at the smallest value $Q=1.2~$GeV displayed in fig.~\ref{fig:AdlerQsqdep}, this uncertainty becomes comparable to the one from $\mu$ and $t$ variation. 

We observe that the reduction in uncertainty due to gradient-flow subtraction persists in the entire energy range from $1.2\,\mbox{GeV}$ to $3\,\mbox{GeV}$ shown.  Roughly speaking, the gradient-flow OPE scheme allows us to predict the Adler function in the low-$Q$ regime at almost 500~MeV smaller values at commensurate errors. If the subtraction term, that is the coefficient $e_3$, was known also to fourth order, the uncertainty would be further reduced. At this point, the region of applicability of the method is presumably limited by higher-dimension condensates, which mandates an analysis of the convergence of the OPE (and in case of $\tau$ decay discussed below, parton-hadron duality violation). 

\subsubsection{Uncertainty from $\alpha_s$}

Before closing this section, we mention the uncertainty
due to the strong coupling value. The current PDG value of $\alpha_s(M_Z)=0.1180(9)$ 
in the $\overline{\rm MS}$ five-flavour scheme corresponds to $\alpha_s(m_{\tau})=0.314(7)$ in the three-flavour scheme.
Since the uncertainty due to this parameter input is independent of the effectiveness of the gradient-flow subtraction method, we neglected it so far. 
However, let us briefly examine the associated uncertainty.
If we look at the GF OPE with $Q=m_{\tau}$ at N$^3$LO, we obtain
\be
\Delta_D(m_{\tau}^2)=0.1358\pm 0.0022 (\alpha_s)  \quad{} (\text{N$^3$LO}).
\ee
This uncertainty is almost independent 
of the perturbative order, and comparable to 
the total uncertainty in the corresponding 
result~\eqref{AdlerMtauGFOPE}.
Remarkably, if we go to the next order (N$^4$LO),
the $\alpha_s$ uncertainty exceeds the total systematic uncertainty of 
the gradient-flow OPE calculation.
This implies that with the GF OPE method the Adler function 
has the potential to contribute to a more precise determination 
of $\alpha_s$.


\section{Hadronic tau decay width}
\label{sec:tau}

\subsection{Preliminaries}

The hadronic tau decay width is one of the most important 
observables where the Adler function plays a central role.
The ratio
\be
R_{\tau}=\frac{\Gamma(\tau^{-} \to {\rm hadrons}+\nu_{\tau})}{\Gamma(\tau^{-} \to e^{-} \bar{\nu}_e \nu_{\tau})}
\ee
can be decomposed as
\be
R_{\tau}=R_{V}+R_{A}+R_{S}
\ee
where $R_{V}$ and $R_{A}$ represent the contributions
from the vector current correlator $J^V_{\mu}=\bar{u} \gamma_{\mu} d$ 
and the axial current correlator $J^A_{\mu}=\bar{u} \gamma_{\mu} \gamma_5 d$, respectively. $R_{S}$ originates from the Cabibbo-suppressed $\bar{u} s$-quark currents. We focus on the sum of the dominant $R_V$ and $R_A$ contributions. They are written as \cite{Braaten:1991qm}
\be
R_{V+A}= N_c S_{\text{EW}} |V_{ud}|^2 \left[1+\delta_{\rm QCD}+\delta'_{\text{EW}}\right] .
\ee  
Here $S_{\text{EW}}=1.0198 \pm 0.0006$ and $\delta_{\rm QCD}$, $\delta'_{\text{EW}}=0.0010 \pm 0.0010$
are the QCD and electroweak corrections, respectively. The QCD 
correction includes the OPE series of power corrections, and is expressed as 
\be
\delta_{\rm QCD} \equiv \delta^{(0)}_{V+A}+\sum_{D \geq 2} \delta_{V+A}^{(D)}\,,
\label{eq:deltatau}
\ee
where $\delta^{(0)}_{V+A}$ denotes the perturbative QCD contribution in the massless-quark limit, and $\delta_{V+A}^{(D)}$ represents finite quark-mass 
and non-perturbative corrections of order $1/Q^D$. When quark masses are neglected, which is a very good approximation for the up- and down-quarks, the vector- and axial-vector contributions differ only from dimension $D\geq 6$. Here we focus on the interplay of the perturbative and dimension-four gluon-condensate terms. We use the same notation \eqref{eq:deltatau} for the gradient-flow OPE expansion. In this case, 
$\delta^{(0)}_{V+A}$ refers to the subtracted, $t$-dependent perturbative contribution, which also contains the $1/t^2$-term of the small-flow time expansion, and $\delta_{V+A}^{(4)}$, the non-perturbative gradient-flowed gluon-condensate contribution. 

The QCD correction to the tau hadronic width can be related to the Adler function by the contour integral \cite{Braaten:1988hc},
\be
\delta_{\rm QCD}=
\frac{1}{2 \pi i} \oint_{|x|=1} \frac{dx}{x} (1-x)^3 (1+x) \Delta_D(-m_{\tau}^2 x)\,, 
\label{eq:deltaQCD}
\ee
where $x=-e^{i \theta}$ with $-\pi < \theta < \pi$.
In the evaluation of the contour integral, there are two conventional choices of the 
renormalization scale:
\begin{align}
&{\text{FOPT}:} \qquad{} \mu^2=m_{\tau}^2 , \non
&{\text{CIPT}:} \qquad{} \mu^2=Q^2=m_{\tau}^2 e^{i \theta} ,
\end{align}
where FOPT and CIPT stand for fixed-order perturbation theory and
contour-improved perturbation theory, respectively.
In FOPT, the running coupling $\alpha_s(\mu)$ remains constant in the contour integration,
whereas the logarithm $L=\log(Q^2/\mu^2)=i \theta$ of
eq.~\eqref{LandLt} depends on $\theta$.
On the other hand, in CIPT,
$\alpha_s(\mu)$ depends on $\theta$, whereas $L$ is a constant
(or zero with the above choice).
This approach thus corresponds to applying the renormalization-group (RG) 
improvement at every point locally on the complex circle. Both  choices are seemingly reasonable. However, it has been long known that these two perturbative approaches do not converge to an identical result as one adds more and more perturbative orders and only FOPT is legitimate in the presence of factorial divergence of the series expansion \cite{Beneke:2008ad}. In ref.~\cite{Beneke:2012vb} it was noted that the problem is related specifically to the $u=2$ renormalon in $C_1(Q^2)$ and gluon-condensate contribution. Ref.~\cite{Gracia:2023qdy} finally revealed the mathematical origin of the problem. Even earlier, it was shown that subtracting the $u=2$ renormalon  largely resolves the discrepancy between the two approaches \cite{Benitez-Rathgeb:2022hfj,Benitez-Rathgeb:2022yqb}. In ref.~\cite{Beneke:2023wkq}, 
we demonstrated how gradient-flow subtraction accomplishes this task without recourse to the Stokes constant of the $u=2$ IR renormalon and explicit reference to large-order behaviour of the perturbative expansion. In the following, we provide further details and an extension of this analysis: We now include $r$ of eq.~\eqref{eq:r} to NNLO and, moreover, the leading non-perturbative gluon condensate correction.

\subsection{Resolving the FOPT/CIPT discrepancy}

We evaluate eq.~\eqref{eq:deltaQCD} using the gradient-flowed OPE~\eqref{eq:newOPE} for the Adler function.  We already clarified how to evaluate the GF OPE of the Adler function for real positive $Q^2$,
and it is straightforward to extend the result to complex-valued $Q^2$. 
The gradient-flow subtraction time $t$ is 
{\em} not complexified. In the results shown below, we therefore adopt
\begin{align}
&{\text{RS FOPT}:} \qquad{} \mu^2=m_{\tau}^2 , \quad{} t=20/(8 m_{\tau}^2)\, , \non
&{\text{RS CIPT}:} \qquad{} \mu^2=Q^2=m_{\tau}^2 e^{i \theta} , \quad{} t=20/(8 m_{\tau}^2)\,,
\end{align}
where ``RS'' stands for ``renormalon-subtracted''. 
From eqs.~\eqref{eq:IRambiguity} and \eqref{eq:Y1u2} it is seen 
that the $u=2$ renormalon properly cancels in the subtracted perturbative Adler function even for complex $Q^2$ and real $t$, irrespective of the choice of $\mu^2$.
Hence, the $u=2$ renormalon should successfully be subtracted in
both, FOPT and CIPT, and the discrepancy is expected to be resolved.

\begin{figure}[t] \centering
   \includegraphics[width=11cm]{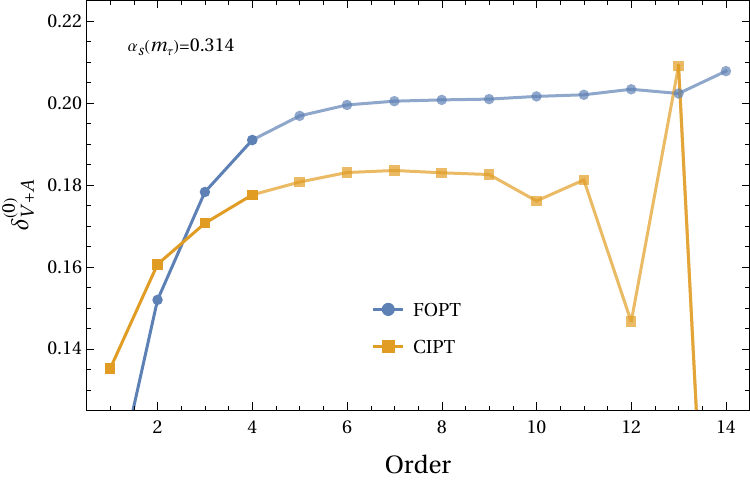}\\[0.2cm]
    \includegraphics[width=11cm]{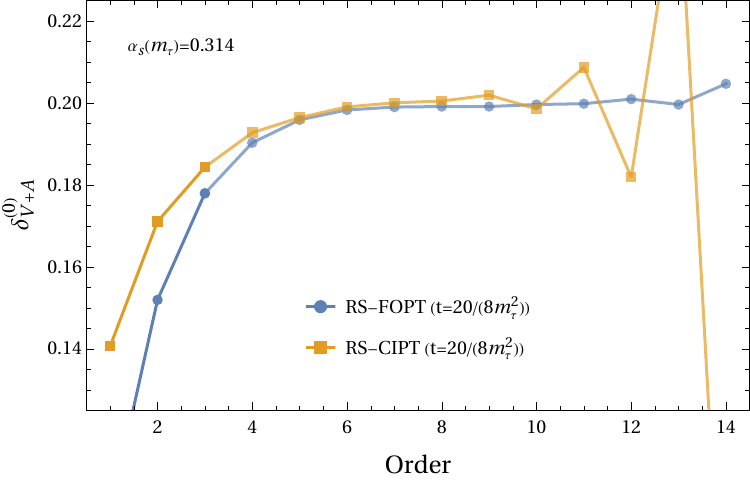}
\caption{Comparison of FOPT and CIPT in the standard perturbative series (upper) and in the gradient-flow subtracted 
series (lower). In the lower figure, the flow time is taken as $8 t m_{\tau}^2=20$. Beyond 4th (3rd) order, estimated 
perturbative coefficients are used and 
indicated by fainter colours. 
\label{fig:FOvsCI}
}
\end{figure}

In the upper panel of fig.~\ref{fig:FOvsCI} the standard
FOPT and CIPT perturbative approximations $\delta_{V+A}^{(0)}$ to the hadronic tau decay width are shown, reproducing \cite{Beneke:2008ad} and illustrating the discrepancy mentioned earlier. On the other hand, in the lower panel,
the evaluation of  $\delta_{V+A}^{(0)}$
employs the subtracted Adler function series
in the two approaches.
The figure demonstrates that the gradient-flow subtraction of the $u=2$ IR renormalon resolves the FOPT/CIPT discrepancy \cite{Beneke:2023wkq}. The 
$t$-dependence of this statement is analyzed in appendix~\ref{app:tdep}. 

By comparing both panels, one notes that the FOPT series hardly changes by the subtraction. This is due to the well-known feature that the (standard) gluon condensate contribution and therefore also the $u=2$ IR renormalon ambiguity to the hadronic tau decay width are suppressed by two powers of $\alpha_s$ relative to the one for the Adler function itself. 
This is related to the integrand of eq.~\eqref{eq:deltaQCD}: due to the weight
function $(1-x)^3 (1+x)$, the gluon-condensate contribution to $\Delta_D(-m_\tau^2 x)$ proportional to $\Lambda^4/Q^4 =\Lambda^4/(m_{\tau}^4 x^2)$ contributes to the contour integral only when $C_{FF}(Q)$ supplies powers of $\ln(-x)$.
This happens first at $\mathcal{O}(\alpha_s^2)$ 
through the appearance of $L$ in eq.~\eqref{eq:CFFexplicit},
i.e.~when $r$ is evaluated in the NNLO approximation, which is the presently highest available order. This is in stark contrast to CIPT, where the cancellation required to engineer this suppression does not take place \cite{Beneke:2012vb}, which already indicates an inconsistency of this approach. The gradient-flow subtraction term rectifies this problem. 

\subsection{Inclusion of the gluon condensate}

This suppression applies equally to the gradient-flowed gluon condensate due to the same leading $Q$ dependence. Compared to the Adler function (see fig.~\ref{fig:tdep}), we expect a minor effect. In the gradient-flowed OPE, the gluon condensate adds not only the leading non-perturbative correction but also serves to cancel the $t$-dependence of the subtracted perturbative contribution.

\begin{figure}[t] 
\begin{center}
\includegraphics[width=11cm]{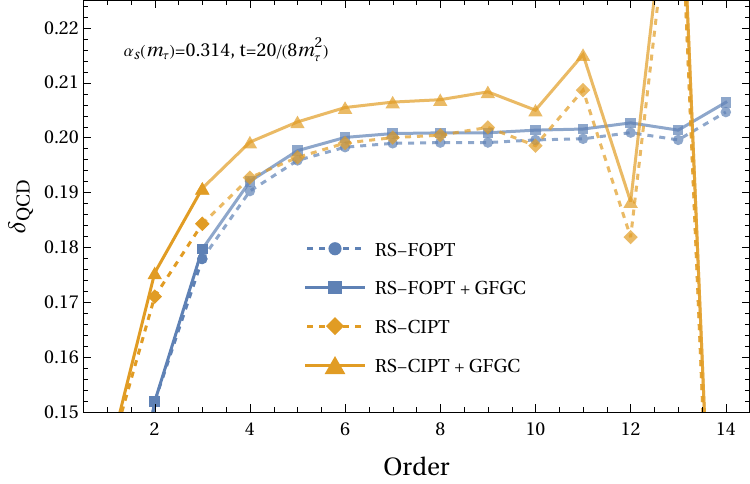}
\end{center}
\caption{ 
$\delta_{\rm QCD}$ based on the GF OPE,
showing the FO (blue) and CI (light orange) approaches. Dashed lines refer to the (subtracted) perturbative approximations, to which the non-perturbative gluon condensate is added (solid lines). Beyond the third order, estimated 
perturbative coefficients are used 
as indicated by the fainter colours.}
\label{fig:tauGFOPEFOvsCI} 
\end{figure}

In fig.~\ref{fig:tauGFOPEFOvsCI}, we show the 
results for the hadronic tau decay width including the gradient-flowed gluon-condensate contribution.\footnote{The $t$ dependence
is seen to be minor in fig.~\ref{fig:FOvsCItdep} in appendix~\ref{app:tdep}, consistent with the observed suppressed contribution of the gradient-flowed gluon condensate.} Focusing first on FO (blue), the shift of the perturbative approximation (dashed) after adding the GF gluon condensate (solid) is not very significant, as expected. 

However, the figure displays a new problem with the CI result (light orange). While the FO-CI discrepancy was resolved for the perturbative expansion of the unit operator in the gradient-flowed OPE, it reappears when the next term in the OPE is added, though reduced in size, since it is related to the non-perturbative correction.  
We attribute the ``new'' discrepancy 
to IR renormalons in the coefficient function $r$ of the gluon-condensate term.
The Wilson coefficients $C_{FF}(Q^2)$ and
$Y_{FF}(t)$, of which $r$ consists, have IR renormalons at $u=1$,
leading to $\mathcal{O}(\LMS^6)$ ambiguities.
Just as CI was inconsistent with the OPE 
when the standard perturbative series (of the unit operator) was not subtracted, the fact that we did not subtract the series expansion of $r$ reintroduces the inconsistency at the level of the first power correction. We emphasize that this is not a problem of the gradient-flow subtraction proposal, but a consequence of the truncation at dimension-4 and the specific problem with the CIPT for the hadronic tau decay width. In the subsequent section we sketch, how gradient-flow subtraction can be extended to dimension 
six operators, which includes the treatment of the $u=1$ renormalon in $r$. However, fig.~\ref{fig:tauGFOPEFOvsCI} does imply that 
without a treatment of dimension-six terms, 
the inclusion of the gradient-flowed 
gluon condensate can be reasonably done only 
in the FO approach, following the arguments in refs.~\cite{Beneke:2008ad,Gracia:2023qdy},
as only FO is consistent with the OPE when IR renormalons are relevant.


\section{Extension to dimension-six operators}
\label{sec:dim6}

The success of gradient-flow subtraction at the level of the dimension-four power correction merits the consideration of higher-dimension operators. One of virtues of eq.~\eqref{eq:newOPE} is that it can easily be extended to include higher-dimension operators, thereby automatically subtracting  sub-leading IR renormalons. 
This is all the more important as the gain in precision allows to go to smaller $Q^2$ (as long as the $1/Q^2$-expansion is valid). 
As a matter of principle, replacing the standard ambiguous condensates by the gradient-flow regularized counterparts  automatically eliminates all IR renormalons in the short-distance coefficients, up to the order in the $1/Q^2$ expansion considered. 
This is because ambiguous condensates which are responsible for cancelling ambiguities in short-distance coefficients are all removed. 

At dimension six, in the massless quark limit, the basis of gauge-invariant, colour-singlet local operators consists of the three-gluon operator 
$g_s^3 f_{ABC} \,G_{\mu\nu}^A G_\rho^{\nu\,B} 
G^{\rho\mu\,C}$ and a large number of four-quark operators with 
field content $(\bar q q) (\bar{q} q)$ and various flavour, spin- and colour structures. Their explicit form and one-loop renormalization can be found in ref.~\cite{Boito:2015joa}. In phenomenological studies \cite{Braaten:1991qm}, the vacuum saturation approximation is often employed. 

Here our aim is to discuss the structure of the gradient-flow subtracted OPE when higher-order power corrections beyond the leading gluon-condensate term are included. Leaving a full treatment of the QCD case to future work, we adopt a highly simplified model with only a single dimension-6 four-quark operator. 
The standard OPE of the Adler function with short-distance coefficients in the $\overline{\rm MS}$ scheme is now written as 
\be
D(Q^2) = \frac{N_c}{12\pi^2} \,\left(C_1(Q^2) 
+\frac{C_{FF}(Q^2)}{Q^4} \langle \frac{\alpha_s}{\pi} F^2 
\rangle 
+\frac{C_{4q}(Q^2)}{Q^6} \langle (\bar{q}q)^2 \rangle
+\mathcal{O}(\LQCD^8/Q^8) \right).
\label{eq:AdlerOPE6}
\ee
The standard four-quark condensate $\langle (\bar{q}q)^2 \rangle$ is also considered to be ambiguous like the standard gluon condensate. In this expression the coefficient function $C_1$ of the unit operator contains IR renormalon divergence, which causes ambiguities that must be cancelled by ambiguities of both condensates. The $u=2$ renormalon singularity is related to the gluon condensate, the next $u=3$ one to the four-quark operator. Once dimension-six operators are included, one can (and must) also deal with the $u=1$ renormalon of the gluon-condensate coefficient function $C_{FF}$. This causes a $\LMS^2/Q^2$ ambiguity in $C_{FF}$, which must also cancel with the ambiguous standard four-quark condensate. These cancellations become automatic when the standard gluon and four-quark condensate are replaced by the gradient-flowed ones.

To extend the gradient-flowed OPE to dimension-six level, we additionally consider the
gradient-flowed counterpart of the dimension-six operator $(\bar{q}q)^2$, which  
will be denoted as $(\bar{\chi} \chi)^2 (t)$.
It is known that flowed quark condensates require wave-function renormalization \cite{Luscher:2013cpa}. Although we do not specify the scheme here,\footnote{The ringed scheme \cite{Makino:2014taa} is one such scheme.} 
$ (\bar{\chi} \chi)^2 (t)$ should be understood as a finite quantity which is accessible through lattice QCD. The small flow-time expansion of this quantity and the action density to
dimension six are given by
\begin{eqnarray}
&&\frac{E(t)}{\pi^2}=\frac{Y_1(t)}{t^2}+Y_{FF}(t) \,\langle \frac{\alpha_s}{\pi} F^2\rangle+
t \,Y_{4q}(t)\, \langle (\bar{q}q)^2 \rangle +\mathcal{O}(t^2 \LQCD^8)\,,
\label{eq:smallflowexp6E}
\\[0.2cm]
&&\langle \left( \bar{\chi} \chi \right)^2 \rangle(t) = 
\frac{X_1(t)}{t^3}+\frac{X_{FF}(t)}{t} \langle \frac{\alpha_s}{\pi} F^2 \rangle + X_{4q}(t) \, \langle (\bar{q}q)^2 \rangle +\mathcal{O}(t \LQCD^8)\,.
\label{eq:smallflowexp6chi}
\end{eqnarray}
The short-distance coefficients $Y_i(t)$, $X_i(t)$ ($i=1, FF, 4q$) have the same
IR renormalon structure as the corresponding $C_i$ and the cancellation of ambiguities with the standard condensates in the $\overline{\rm MS}$ scheme small flow-time expansion must follow the same pattern as described for $C_i$ above. 
From now on it will be understood that the coefficient functions $C_i$ depend on $Q^2$, while the $X_i$, $Y_i$ depend on $t$, and the corresponding arguments will be omitted to facilitate the reading of equations. 

To eliminate the ambiguous condensates from the OPE of the Adler function, as before, we solve the previous two equations for $\langle \frac{\alpha_s}{\pi} F^2 \rangle$ and $\langle (\bar{q}q)^2 \rangle$ to obtain 
\begin{align}
\langle \frac{\alpha_s}{\pi} F^2 \rangle
=\frac{1}{t^2} \frac{1}{\Delta(t)} \left[
X_{4q} \left(\frac{t^2 E(t)}{\pi^2}-Y_1\right)
-Y_{4q}\, (t^3 \langle \left( \bar{\chi} \chi \right)^2 \rangle(t)-X_1 
\right]+\mathcal{O}(t^2 \LQCD^8)\, , 
\label{eq:GCdim6}
\end{align}
\begin{align}
\langle (\bar{q} q)^2 \rangle
=\frac{1}{t^3} \frac{1}{\Delta(t)} \left[- 
X_{FF} \left(\frac{t^2 E(t)}{\pi^2}-Y_1 \right)
+Y_{FF}(t) \,(t^3 \langle \left( \bar{\chi} \chi \right)^2 \rangle(t)-X_1 ) 
\right]+\mathcal{O}(t \LQCD^8)\, , 
\label{eq:qbarqdim6}
\end{align}
where $\Delta(t)$ is given by
\begin{equation}
    \Delta(t) = Y_{FF}X_{4q}- Y_{4q}X_{FF}\,.
\end{equation}
Apparently, the gluon condensate \eqref{eq:GCdim6} has a different form from 
the earlier eq.~\eqref{eq:GCdim4}.
This is simply an algebraic consequence:
it is expressed using two gradient-flowed observables instead of one.
One can of course check that eq.~\eqref{eq:GCdim6} reduces to 
the previous expression for $\langle \frac{\alpha_s}{\pi} F^2 \rangle$
once the small flow time expansions are applied.
The more important point is that, 
the gluon condensate has now been 
expressed more precisely---compare $\mathcal{O}(...)$ in eqs.~\eqref{eq:GCdim4} and \eqref{eq:GCdim6}.
In other words, making use of two gradient-flow observables helps improve the approximation.
From the above results, we obtain the gradient-flowed OPE of the Adler function in the form 
\begin{eqnarray}
D(Q^2)&=&\frac{N_c}{12\pi^2} \,\Bigg[ 
\,\bigg(C_1-\frac{C_{FF}}{t^2 Q^4} 
\frac{Y_1 X_{4q}-Y_{4q} X_1}{\Delta} 
-\frac{C_{4q}}{t^3 Q^6} 
\frac{Y_{FF} X_{1}-Y_{1} X_{FF}}{\Delta}  \, \bigg)
\nonumber\\
&&+\,
\frac{C_{FF} X_{4q}-\frac{C_{4q} X_{FF}}{t Q^2} }{\Delta}
\frac{1}{Q^4}  \frac{E(t)}{\pi^2}
\nonumber\\
&&
+\,\frac{C_{4q} Y_{FF}-t Q^2 \,C_{FF} Y_{4q} }{\Delta}
\frac{1}{Q^6}\, \langle \left( \bar{\chi} \chi \right)^2 \rangle(t)
\nonumber\\
&&+\,\mathcal{O}(t^2 \LQCD^8/Q^4,t \LQCD^8/Q^6, \LQCD^8/Q^8) \,\Bigg] .
\label{eq:dim6newOPE}
\end{eqnarray}

One can show that this formula is free
from IR renormalons up to $\mathcal{O}(\LQCD^6)$ as follows.
The initial task is to understand the ambiguities of each quantity.
To this end, we first go back to the original OPEs ~\eqref{eq:AdlerOPE6}, \eqref{eq:smallflowexp6E}, \eqref{eq:smallflowexp6chi} 
and formally
decompose the short-distance coefficients  into
their ambiguous (imaginary) and unambiguous (real) part of their Borel integrals. 
For instance, 
\begin{align}
C_1
&=C_{1,\rm PV} \pm i \delta C_1|_{u=2} \pm i \delta C_1|_{u=3} \, , \non
C_{FF}
&=C_{FF,\rm PV} \pm i \delta C_{FF}|_{u=1} \, , \non
C_{4q}
&=C_{4q,\rm PV} \, . \label{eq:decompWilson}
\end{align}
PV represents the principal value integral of the Borel integral and thus its real part, while $\delta ...|_{u=...}$ corresponds to imaginary part
of the Borel integral. The value of $u$ controls the scaling $\delta C_{...}(Q^2)|_{u=v} \approx \LQCD^{2v}/Q^{2v}$. Note that our consideration is restricted to the level of $\mathcal{O}(\LQCD^6)$, and renormalon ambiguities are considered up to this order. An analogous decomposition applies to the other short-distance coefficients $Y_i$ and $X_i$. Likewise, 
the condensates are decomposed into their real ``principal value'' and  ambiguous imaginary part:
\begin{align}
\langle \frac{\alpha_s}{\pi} F^2 \rangle
&=\langle \frac{\alpha_s}{\pi} F^2 \rangle_{\rm PV} \pm i \delta \langle \frac{\alpha_s}{\pi} F^2 \rangle \, , \non
\langle (\bar{q}q)^2 \rangle
&=\langle (\bar{q}q)^2 \rangle_{\rm PV}
\pm i \delta \langle (\bar{q}q)^2 \rangle \, .
\end{align}
Then, for instance for the Adler function, one obtains, 
arranging in powers of $\LQCD/Q$:
\begin{eqnarray}
D(Q^2)
&=&\frac{N_c}{12 \pi^2} 
\bigg[
C_{1,\rm PV}+\frac{C_{FF, \rm PV}}{Q^4} \langle \frac{\alpha_s}{\pi} F^2 \rangle_{\rm PV}+
\left(\frac{C_{4q}}{Q^6}\, \langle (\bar{q}q )^2 \rangle_{\rm PV}
-\frac{\delta C_{FF}|_{u=1}}{Q^4} \delta \langle \frac{\alpha_s}{\pi} F^2 \rangle \right)
\bigg] \non
&&\pm i\, \frac{N_c}{12 \pi^2}
\bigg[
\left(
\delta C_1|_{u=2}+\frac{C_{FF, \rm PV}}{Q^4} \delta \langle \frac{\alpha_s}{\pi} F^2 \rangle
\right) 
\non
&&\hspace*{1.35cm}
+\left(
\delta C_1|_{u=3}
+\frac{\delta C_{FF}|_{u=1}}{Q^4} 
\langle \frac{\alpha_s}{\pi} F^2 \rangle_{\rm PV}
+C_{4q,\rm PV} \,\delta \langle (\bar{q}q )^2 \rangle
\right)
\bigg] \non
&&+\mathcal{O}(\LQCD^8/Q^8) \, . 
\label{eq:decomAlder}
\end{eqnarray}
The first line corresponds to the real value of the Adler function,
while the following lines give relations that the ambiguous quantities should satisfy so that the ambiguous imaginary part is zero. This implies 
\begin{align}
0&=\delta C_1|_{u=2}+\frac{C_{FF,\rm PV}}{Q^4} \delta \langle \frac{\alpha_s}{\pi} F^2 \rangle \, , \non
0&=\delta C_1|_{u=3}
+\frac{\delta C_{FF}|_{u=1}}{Q^4} 
\langle \frac{\alpha_s}{\pi} F^2 \rangle_{\rm PV}
+C_{4q,\rm PV} \,\delta \langle (\bar{q}q )^2 \rangle \, . \label{eq:renomcancelrel}
\end{align}
Analogous results can be obtained for
$E(t)$ and $\langle (\bar{\chi} \chi)^2 \rangle (t)$ with the $C_i$ replaced by the corresponding $Y_i$ and $X_i$, respectively.
At this point we go back to the OPE~\eqref{eq:dim6newOPE}.
We apply the decompositions such as eq.~\eqref{eq:decompWilson}
in eq.~\eqref{eq:dim6newOPE}.
Simultaneously, we replace
$E(t)$ and $\langle (\bar{\chi} \chi)^2 \rangle (t)$ in eq.~\eqref{eq:dim6newOPE} with their ``real values'' 
in the sense explained below eq.~\eqref{eq:decomAlder}.
Expanding in $\LQCD$ and 
making use of the renormalon cancellation relations such as eq.~\eqref{eq:renomcancelrel},
one shows that the imaginary parts vanish, implying that
eq.~\eqref{eq:dim6newOPE} is free of 
renormalons of up to $\mathcal{O}(\LQCD^6)$.
One may say that the new OPE~\eqref{eq:dim6newOPE} 
inherits the property that the original OPEs are in total unambiguous, but makes this explicit, since the gradient-flow condensates are already unambiguous.

We make some comments.
First, eq.~\eqref{eq:dim6newOPE} can be regarded as an extension 
in the sense that it reduces to eq.~\eqref{eq:newOPE} in the formal limit where we neglect $\langle \left( \bar{\chi} \chi \right)^2 \rangle(t)$ (and the associated Wilson coefficients $X_i$).
For instance, the first two terms in the first line of eq.~\eqref{eq:dim6newOPE} can be written as
\be
C_1(Q^2)-\frac{1}{t^2 Q^4} r(Q^2,t) Y_1\left[
\frac{1-\frac{Y_{4q} X_1}{X_{4q} Y_1}}{1-\frac{Y_{4q} X_{FF}}{X_{4q} Y_{FF}}}\right] \, .
\ee
Compare this with (the first line of) eq.~\eqref{eq:newOPE}.
Second, the renormalon cancellation structure within the formula is fairly complicated.
Although the $\mathcal{O}(\LQCD^4)$ ambiguities, including the $u=2$ renormalon in the unit operator coefficients, are eliminated within
the terms inside the round parentheses of the first line of eq.~\eqref{eq:dim6newOPE}, the cancellation of other ambiguities requires the inclusion of the 
(non-perturbative) $E(t)$ and $\langle \left( \bar{\chi} \chi \right)^2 \rangle(t)$ terms. 
This is related to the fact that 
the relations of eq.~\eqref{eq:renomcancelrel} give only very general  constraints.

However, based on the following observation, 
stronger constraints than eq.~\eqref{eq:renomcancelrel} could be formulated. 
With a gauge-invariant UV cut-off (such as the gradient-flow time), the power UV divergences of the dimension-six operator can be decomposed as
\be
(\bar{q} q)^2 \sim h(\alpha_s) \Lambda_{\rm UV}^6 \,1
+g(\alpha_s) \Lambda_{\rm UV}^2 \frac{\alpha_s}{\pi} F^2+\dots \,.
\label{eq:qqUVdivs}
\ee
Then the $u=1$ renormalon and the $u=3$ renormalon can be related to the quadratic and sixth-power UV divergence of
the dimension-six operator individually:
\begin{equation}
\delta C_{1}|_{u=3} = H C_{4q} \frac{\LMS^6}{Q^6} \,,\qquad
\delta C_{FF}|_{u=1} = G C_{4q} \frac{\LMS^2}{Q^2} \,.
\end{equation}
This complements the relation 
\begin{equation}
\delta C_{1}|_{u=2} = K C_{FF} \frac{\LMS^4}{Q^4} 
\end{equation}
from eq.~\eqref{eq:IRambiguity}. Identical relations hold when $C_i$ on the left and right are substituted by $Y_i$ or $X_i$, with the {\em same} constants $K,H,G$ as above, since eq.~\eqref{eq:qqUVdivs} is an operator statement. 
This means that the second equation of
eq.~\eqref{eq:renomcancelrel} actually contains two independent constraints.

When the above relations hold, the renormalon cancellation structure becomes drastically simpler, and the coefficient functions of the unit operator, $E(t)$ and $\langle (\bar{\chi}\chi)^2\rangle(t)$ of the gradient-flowed OPE are individually free from renormalon divergence (up to terms related to higher dimension $\LQCD^8$). To see this explicitly, it is convenient to rewrite 
eq.~\eqref{eq:dim6newOPE} as 
\begin{eqnarray}
D(Q^2)&=&\frac{N_c}{12\pi^2} \,\Bigg\{
C_1-\frac{1}{t^2 Q^4} \left[C_{FF}-\frac{1}{t Q^2} \frac{C_{4q}}{X_{4q}}X_{FF}\right] \frac{Y_1}{Y_{FF}}-\,\frac{1}{t^3 Q^6} \frac{C_{4q}}{X_{4q}} X_1
\nonumber\\
&&\hspace*{1.35cm}+\frac{1}{t^2 Q^4} \,\frac{X_1-\frac{X_{FF}}{Y_{FF}}Y_1}{Y_{FF}-\frac{Y_{4q}}{X_{4q}}X_{FF} }\frac{Y_{4q}}{X_{4q}}
\left[C_{FF}-\frac{1}{t Q^2}  \frac{C_{4q}}{X_{4q}}X_{FF}\right]
\nonumber\\
&&+\,
\frac{C_{FF}-\frac{1}{t Q^2}\frac{C_{4q}}{X_{4q}} X_{FF}}{Y_{FF}-\frac{Y_{4q}}{X_{4q}} X_{FF}}
\frac{E(t)}{\pi^2}
\nonumber\\[0.1cm]
&&
+\,\frac{C_{4q} Y_{FF}-t Q^2 \,C_{FF} Y_{4q} }{\Delta}
\frac{1}{Q^6}\, \langle \left( \bar{\chi} \chi \right)^2 \rangle(t)
\nonumber\\
&&+\,\mathcal{O}(t^2 \LQCD^8/Q^4,t \LQCD^8/Q^6, \LQCD^8/Q^8) \,\Bigg\}\,.
\label{eq:dim6newOPErearranged}
\end{eqnarray}
The first two lines represent the gradient-flow subtracted unit-operator coefficient, the third (fourth) line the one of the gradient-flowed 
gluon (four-quark) condensate.

\paragraph{Cancellation of the $u=2$ renormalon} This is present only in $C_1$, $Y_1$, $X_1$ and the unit-operator coefficient. Use 
\begin{equation}
\delta C_1=K C_{FF} \frac{\LMS^4}{Q^4} \,,\quad
\delta Y_1=K Y_{FF} \LMS^4 t^2\,,\quad
\delta X_1=K X_{FF} \LMS^4 t^2\,.
\end{equation}
Then from the first line we get 
\begin{equation}
K  \frac{\LMS^4}{Q^4} \left[C_{FF}-C_{FF} \frac{Y_{FF}}{Y_{FF}}\right] +
K\frac{\LMS^4}{t Q^6} C_{4q}\left[\frac{X_{FF}}{X_{4q}}\frac{Y_{FF}}{Y_{FF}}-\frac{X_{FF}}{X_{4q}}\right] = 0
\end{equation}
while in the second line the cancellation takes place already within the numerator
\begin{equation}
X_1-\frac{X_{FF}}{Y_{FF}} Y_1\,.
\end{equation}

\paragraph{Cancellation of the $u=1$ renormalon} This is present only in $C_{FF}$, $Y_{FF}$, $X_{FF}$. Here one needs to consider the coefficient of the unit operator and of the action density $E(t)/\pi^2$. Use 
\begin{equation}
\delta C_{FF} = G C_{4q} \frac{\LMS^2}{Q^2} \,,\quad
\delta Y_{FF}=G Y_{4q} \LMS^2 t\,,\quad
\delta X_{FF}=G X_{4q} \LMS^2 t\,.
\end{equation}
From these relations the cancellation in the coefficient of $E(t)/\pi^2$ in the third line of eq.~\eqref{eq:dim6newOPErearranged} is immediate. Concerning the unit-operator coefficient, the square brackets in the first and second line already contain the cancellation of the ambiguity from $C_{FF}$. Also the denominator in the second line has a manifest cancellation. What remains from the first and second line is 
\begin{eqnarray}
&&-\frac{1}{t^2 Q^4} \left[C_{FF}-\frac{1}{t Q^2} \frac{C_{4q}}{X_{4q}}X_{FF}\right] \times G\frac{\LMS^2}{t Q^4} 
\nonumber\\
&&
\times\,\left[-\frac{Y_1 Y_{4q}}{Y_{FF}^2}-\frac{-\frac{X_{4q}}{Y_{FF}}Y_1+\frac{Y_{4q}X_{FF} Y_1}{Y_{FF}^2}}{Y_{FF}-X_{FF} \frac{Y_{4q}}{X_{4q}}}\frac{Y_{4q}}{X_{4q}}\right] =0\,,
\end{eqnarray}
since the square bracket in the second line of this expression is zero.

\paragraph{Cancellation of the $u=3$ renormalon} This is present only in $C_{1}$, $Y_{1}$, $X_{1}$ and the unit-operator coefficient. Use 
\begin{equation}
\delta C_{1} = H C_{4q} \frac{\LMS^6}{Q^6} \,,\quad
\delta Y_{1}=H Y_{4q} \LMS^6 t^3\,,\quad
\delta X_{1}=H X_{4q} \LMS^6 t^3\,.
\end{equation}    
There is a cancellation between the first and the last term in the first line. The second term in the first line and the second line cancel upon noticing that 
\begin{eqnarray}
&&\delta \left[\frac{Y_1}{Y_{FF}}\right] = H \LMS^6 t^3\frac{Y_{4q}}{Y_{FF}}
\nonumber\\
&&\delta\left[\frac{X_1-\frac{X_{FF}}{Y_{FF}}Y_1}{Y_{FF}-X_{FF} \frac{Y_{4q}}{X_{4q}}}\frac{Y_{4q}}{X_{4q}}\right] = H \Lambda^6 t^3 
\frac{X_{4q}-\frac{X_{FF}}{Y_{FF}}Y_{4q}}{Y_{FF} X_{4q}-X_{FF} Y_{4q}} Y_{4q} = 
 H \LMS^6 t^3\frac{Y_{4q}}{Y_{FF}}\qquad
\end{eqnarray}
\noindent
This concludes the demonstration of the cancellation. We will leave a more comprehensive and realistic treatment of the gradient-flowed OPE of the Adler function at dimension-six  for our future work, but emphasize  
that independent of the details, the renormalon cancellation itself is 
always guaranteed when the standard condensates are replaced by their gradient-flowed counterparts.

\section{Conclusion}
\label{sec:conclusion}

The results of this investigation demonstrate that expressing the local operators in the OPE in terms of their gradient-flowed counterparts efficiently subtracts the IR renormalon factorial divergence of the $\overline{\rm MS}$-scheme short-distance coefficients, while yielding unambiguous non-perturbative operator matrix elements, which can be computed on the lattice and extrapolated to the continuum limits, since their power divergences are regulated by the gradient-flow time $t$. To the best of our knowledge, this results for the first time in a practical and easy-to-implement scheme to consistently combine perturbative series in the strong coupling with exponentially small terms (power corrections in the momentum variable). On top of that, our framework in principle enables the subtraction of {\it all} IR renormalons for the first time by expressing all condensates in terms of their gradient-flowed counterparts. 
The key advantage of the gradient flow resides in manifest gauge invariance, and the fact that it decouples the continuum limit $a\to 0$ of lattice simulations from the UV cut-off $1/\sqrt{2 t}$ of the renormalized matrix elements. The {\em gradient-flowed} OPE therefore realizes a Wilsonian-type hard cut-off, which is essential to eliminate the IR renormalons in the perturbative short-distance coefficients. 

The gradient-flow subtraction does not eliminate all factorial divergence from the perturbative series. However, the ones remaining are either highly suppressed (instanton-anti-instanton singularities) or sign-alternating (UV renormalons). While the UV renormalon divergences are technically as (or even more) important as the IR ones as far as the asymptotic strength  of their factorial growth is concerned, their overall normalization is often suppressed, such that their effect can be ignored at the relevant intermediate orders. If necessary, they could be treated by conformal mapping techniques \cite{Altarelli:1994vz,Caprini:1998wg} (and references therein), and hence they do not restrict the applicability of the gradient-flow subtraction.

We exemplified these features for the Adler function, focusing on the low-energy region $Q \approx (1-3)\,$GeV. At the lower end of this interval, the series of exactly known terms to order $\alpha_s^4$ already reaches the regime where individual terms begin to grow due to the leading IR renormalon singularity. In contrast, after the gradient-flow rearrangement of the OPE and the corresponding  subtraction of the perturbative series,  the perturbative Adler function stays almost constant from order $\alpha_s^2$ (fig.~\ref{fig:fixedorderadler}), and does not display large corrections any more. The leading IR renormalon is related exclusively to the gluon condensate, which in the gradient-flow OPE is simply the action density. By using existing lattice data, we achieve non-perturbative $\LMS^4/Q^4$ accuracy for the Adler function, and a significant reduction of the theoretical uncertainty, allowing one to extend the prediction to lower $Q$ (fig.~\ref{fig:AdlerQsqdep}). These results rely in part on an extrapolation of the lattice data to larger flow times $t$ than presently available. It is highly desirable to  cover flow times up to 3~GeV${}^{-2}$ in lattice simulations extrapolated to the continuum limit.  To check our conclusions, we verified that the combination of the perturbative and non-perturbative contributions are almost independent of the factorization scale $1/\sqrt{2 t}$, despite their (expected) individual large dependence on it (fig.~\ref{fig:tdep}). 

As an added benefit, the gradient-flow subtraction also solves the long-standing problem \cite{Beneke:2008ad} of the inconsistency of the so-called fixed-order and contour-improved approach to computing the inclusive hadronic tau-lepton decay width in favour of the fixed-order approach, as already shown in \cite{Beneke:2023wkq}. This has implications for the determination of the strong coupling from this process. The gain in accuracy for the Adler function itself also offers a new perspective on $\alpha_s$ extractions from it at low $Q$, which we plan to investigate. Furthermore, with better control over power corrections, studies of duality violations become more meaningful. One can also contemplate revisiting results on hadronic quantities from SVZ sum rules \cite{Shifman:1978by} and the strange-quark mass, now with a rigorous definition of the gluon condensate. Last but not least, we envisage applications of the method to other areas such as the heavy-quark expansion, since the gradient-flow definition is  not restricted to vacuum matrix elements.

\subsubsection*{Acknowledgements}
We thank F.~Lange, A.~Saez Gonzalvo and A.~Ramos for helpful discussion and correspondence. This research was supported by the Excellence Cluster ORIGINS, which is funded by the Deutsche Forschungsgemeinschaft (DFG, German Research Foundation) under Germany's Excellence Strategy --- EXC 2094 --- 390783311. HT thanks the Excellence Cluster ORIGINS and the T31 group at Technical University of Munich for their hospitality while part of this work was performed. He also thanks R. Kitano for financial support. The work of HT was supported 
by JSPS KAKENHI Grant Numbers JP18H05542, JP19K14711 and JP23K13110. HT is  Yukawa Research Fellow supported by Yukawa Memorial Foundation.

\appendix

\section{Perturbative series}
\label{app:series}

\subsection{Beta function}

The beta function coefficients $\beta_k$, defined in eq.~\eqref{eq:betafn}, are given by \cite{Baikov:2016tgj}
\begin{eqnarray}
\beta_0 &=&
-\frac{1}{4\pi} \lp 11-\frac{2}{3} n_f \rp \, , \\[-0.2cm] \nonumber \\[-0.1cm]
\beta_1 &=&
-\frac{1}{(4\pi)^2} \lp 102-\frac{38}{3} n_f \rp \, , \\[-0.2cm] \nonumber \\
\beta_2 &=&
-\frac{1}{(4\pi)^3} \lp \frac{2857}{2}-\frac{5033}{18} n_f+\frac{325}{54} n_f^2 \rp \, , \\[-0.2cm] \nonumber \\
\beta_3 &=&
-\frac{1}{(4\pi)^4} \bigg[
\frac{149753}{6}+3564\, \zeta_3
+\lp -\frac{1078361}{162}-\frac{6508}{27} \zeta_3 \rp n_f \nonumber \\
&&\qquad{}\qquad{}+\lp \frac{50065}{162}+\frac{6472}{81} \, \zeta_3 \rp n_f^2
+\frac{1093}{729} n_f^3
\bigg]
\, , \\[-0.2cm] \nonumber \\
\beta_4 &=&
-\frac{1}{(4\pi)^5} \bigg[\frac{8157455}{16}+\frac{621885}{2} \, \zeta_3-\frac{88209}{2} \, \zeta_4-288090 \, \zeta_5 \nonumber\\
&&\qquad{}\qquad{}+\lp -\frac{336460813}{1944}-\frac{4811164}{81} \, \zeta_3+\frac{33935}{6} \, \zeta_4+\frac{1358995}{27} \, \zeta_5\rp n_f \nonumber\\
&&\qquad{}\qquad{}+\lp \frac{25960913}{1944}+\frac{698531}{81} \zeta_3 -\frac{10526}{9} \zeta_4-\frac{381760}{81} \, \zeta_5 \rp n_f^2 \nonumber\\
&&\qquad{}\qquad{}+\lp -\frac{630559}{5832}-\frac{48722}{243} \zeta_3+\frac{1618}{27} \zeta_4+\frac{460}{9} \zeta_5 \rp n_f^3 \nonumber\\
&&\qquad{}\qquad{}+\lp \frac{1205}{2916}-\frac{152}{81} \, \zeta_3 \rp n_f^4 \bigg] \, .
\end{eqnarray}

\subsection{Adler function}

The perturbative part of the Adler function (divided by $N_c/(12\pi^2)$ as in eq.~\eqref{eq:OPE}) is expanded as 
\be
C_1(Q^2)=\sum_{n=0}^{\infty} a_{\mu}^n \sum_{k=1}^{n+1} k c_{n,k} L^{k-1}\,.
\ee
 The coefficients $c_{n,k}$ refer to the perturbative expansion in $\alpha_s$ and $L$ of the correlation function $\Pi(Q^2)$ defined eq.~\eqref{eq:PI}. The $c_{n,k}$, $k>1$ are by RG invariance from $c_{n,1}= c_n(t=\infty)\equiv c_n$. 
 With $N_c=3$ and general $n_f$, the 
analytic results are $c_0=c_1=1$, and 
\begin{eqnarray}
c_2 &=&
\frac{365}{24}-11 \, \zeta_3 +\lp-\frac{11}{12}+\frac{2}{3} \, \zeta_3 \rp n_f
\, , \\[0.2cm]
c_3 &=&
\frac{87029}{288}-\frac{1103}{4} \, \zeta_3+\frac{275}{6} \, \zeta_5 \nonumber\\
&&+\lp-\frac{7847}{216}+\frac{262}{9} \, \zeta_3-\frac{25}{9} \, \zeta_5 \rp n_f 
+\lp\frac{151}{162}-\frac{19}{27} \, \zeta_3 \rp n_f^2 
\, , \\[0.2cm]
c_4 &=& 
\frac{144939499}{20736}
-\frac{5693495}{864}  \,\zeta_{3}
+\frac{5445}{8}  \,\zeta_3^2
+\frac{65945}{288}  \,\zeta_{5}
-\frac{7315}{48}  \,\zeta_{7}
\nonumber\\
&&+\,\lp
-\frac{13044007}{10368}
+\frac{12205}{12}  \,\zeta_{3}
-55  \,\zeta_3^2
+\frac{29675}{432}  \,\zeta_{5}
+\frac{665}{72}  \,\zeta_{7}
\rp n_f
\nonumber\\
&&+\,\lp
\frac{1045381}{15552}
-\frac{40655}{864}  \,\zeta_{3}
+\frac{5}{6}  \,\zeta_3^2
-\frac{260}{27}  \,\zeta_{5}\rp n_f^2
\nonumber\\
&&+\,\lp
-\frac{6131}{5832}
+\frac{203}{324}  \,\zeta_{3}
+\frac{5}{18}  \,\zeta_{5}\rp  n_f^3\,.
\end{eqnarray}
with the $\mathcal{O}(\alpha_s^4)$ (five-loop) term in the (non-singlet) Adler function from ref.~\cite{Baikov:2008jh}.
When a flavour-singlet current 
such as $\bar{u} \gamma_{\mu} u$
is considered,
one must shift $c_3 \to c_3+\delta c_3$, $c_4 \to c_4+\delta c_4$
with \cite{Baikov:2012zn}
\begin{eqnarray}
\delta c_3 &=&
\frac{55}{216}-\frac{5}{9} \, \zeta_3 \, ,  \\
\delta c_4 &=&
\frac{5795}{576}-\frac{8245}{432} \,\zeta_3-\frac{55}{12} \,\zeta_3^2+\frac{2825}{216} \, \zeta_5 \nonumber \\
&&+\lp -\frac{745}{1296}+\frac{65}{72} \, \zeta_3+\frac{5}{18} \, \zeta_3^2-\frac{25}{36} \, \zeta_5 \rp n_f \, .
\end{eqnarray}

The gluon-condensate Wilson coefficient has been calculated
for the vacuum polarization function $\Pi(Q^2)$ at NNLO \cite{Harlander1998,Bruser:2024zyg}.
In ref.~\cite{Bruser:2024zyg} the current correlator 
\be
3 Q^2 \Pi^{V}(Q^2)
=i \int d^4 x e^{i q x} 
\langle 0| T j^{V, \mu}_{1}(x) j^{V}_{\mu, 2}(0) | 0 \rangle 
\label{eq:PiJ}
\ee
with general quark-flavour composition is considered, 
where
\be
j^V_{i, \mu}=\bar{q} \gamma_{\mu} \rho_i q \,, 
\quad{}  (i=1,2) 
\ee
and $\rho_i$ is an $(n_f \times n_f)$-flavour matrix.
The correlator of $j_{\mu}=\bar{u} \gamma_{\mu} d$ and its Hermitian conjugate,
which we consider in this paper,
can be discussed within this treatment.
The result reads
\be
\Pi^{V}(Q^2)
=\cdots+(c_{1,1}^{V,T} {\rm tr} [\rho_1]{\rm tr} [\rho_2]+c_{1,2}^{V,T} {\rm tr} [\rho_1 \rho_2] ) \langle F^a_{\mu \nu} F^{a \mu \nu} \rangle/Q^4+\cdots
\ee
where $c_{1,1}^{V,T}$ and $c_{1,2}^{V,T}$ are
given in ref.~\cite{Bruser:2024zyg}.
We give $C_{FF}(Q^2)$ using $\rho_i$ to facilitate the extension to other cases: 
\begin{eqnarray}
C_{FF}(Q^2) 
&=&{\rm tr} [\rho_1 \rho_2]\,\frac{\pi^2}{N_c}\,T_F\,
\Bigg\{
4 + a_{\mu} \bigg[ 2 C_A -C_F -\frac{4 \pi \beta_1}{\beta_0} \bigg] \nonumber \\
&&
+a_{\mu}^2\,
\bigg[-\frac{69}{8} C_F^2 +
\lp -\frac{209}{24}+\frac{10}{3} \, \zeta_3 \rp C_A^2  \nonumber\\
&&+\lp \frac{1111}{72} C_F +\lp \frac{1}{6}+\frac{4}{3} \, \zeta_3 \rp n_f T_F -\frac{2 \pi \beta_1}{\beta_0}  \rp C_A \nonumber\\
&&
+\lp \lp -\frac{67}{36}+4 \, \zeta_3  \rp n_f T_F+\frac{\pi \beta_1}{\beta_0} \rp C_F 
+\lp \frac{4 \pi^2 \beta_1^2}{\beta_0^2}-\frac{4 \pi^2 \beta_2}{\beta_0} \rp  \nonumber \\
&&
+\lp C_A^2 -\frac{4}{3} n_f  T_F C_F+\lp \frac{11}{12} C_F -n_f T_F \rp C_A \rp L  \,\,\bigg]
\Bigg\} \nonumber \\
&&+{\rm tr} [\rho_1]{\rm tr} [\rho_2]
\,\frac{\pi^2}{N_c}\,
a_{\mu}^2 \,\bigg[ \frac{44}{3}-32 \, \zeta_3 \bigg]\, \frac{d_{FF}^{(3)}}{N_A}+\mathcal{O}(a_{\mu}^3) \, .
\label{eq:CFFexplicit}
\end{eqnarray}
For the current correlator given by eq.~\eqref{eq:PI}, 
one obtains $C_{FF}(Q^2)$ by setting ${\rm tr} [\rho_1 \rho_2]=1$ and ${\rm tr} [\rho_1]{\rm tr} [\rho_2]=0$.
For the correlator of a flavour-singlet vector current such as $j_{\mu}=\bar{u} \gamma_{\mu} u$, 
${\rm tr} [\rho_1 \rho_2]=1$ and ${\rm tr} [\rho_1]{\rm tr} [\rho_2]=1$. The above expression also holds for correlator of the axial-vector current $j^{A}_{i, \mu}
=\bar{q} \gamma_{\mu} \gamma_5 \rho_i q$ of massless quarks as long as ${\rm tr} [\rho_1]{\rm tr} [\rho_2]=0$, which holds for tau decays. 
The colour factors are given by  
\begin{equation}
C_F=\frac{N_c^2-1}{2 N_c}\,, \quad C_A=N_c\,, 
\quad T_F=1/2\,, 
\quad d_{FF}^{(3)}=\frac{(N_c^2-4)(N_c^2-1)}{16 N_c}\,.
\end{equation}

\subsection{Action density}

For the small flow-time expansion of the action density, the Wilson coefficient of the unit operator, $Y_1$,
is given by 
\be
Y_1(t)=\sum_{n=0}^{\infty} e_n a^{n+1}_\mu\,,
\ee
where \cite{Artz:2019bpr}
\begin{eqnarray}
e_0 &=& \frac{3}{4 \pi^2} \frac{N_A}{8} \, , \\
e_1 &=& \frac{3}{4 \pi^2} \frac{N_A}{8} \frac{1}{4}
\,\Bigg\{ 
\lp \frac{52}{9}+\frac{22}{3} \ln 2-3 \ln 3 \rp C_A-\frac{8}{9} T_F n_f-4\pi \beta_0 L_z
\Bigg\} \, , \\[0.2cm]
e_2 &=& \frac{3}{4 \pi^2} \frac{N_A}{8} \frac{1}{16}\,
\Bigg\{ 27.9786 \,C_A^2-31.5652\,C_A T_F n_f
+\lp -\frac{43}{3}+16 \zeta_3 \rp C_F T_F n_f \nonumber \\
&&+\lp -\frac{80}{81}+\frac{8 \pi^2}{27} \rp T_F^2 n_f^2 -4\pi \beta_0 \left[ \lp \frac{104}{9}+\frac{44}{3} \ln 2-6 \ln 3 \rp C_A-\frac{16}{9} T_F n_f \right] L_z\nonumber \\
&&- \,(4 \pi)^2 \beta_1 L_z +(4\pi)^2 \beta_0^2 L_z^2
\,\Bigg\}\, .
\end{eqnarray}
Here $N_A=N_c^2-1$ and we used 
\be
L_z \equiv \log(2 z)+\gamma_E \quad{} (z=\mu^2 t)\,.
\ee
It is straightforward to rewrite these coefficients in terms of $L_t$ through
\be
L_z=-L_t-\ln 4+\gamma_E \,.
\ee

The gluon-condensate Wilson coefficient in the small flow-time expansion of the action density is given by \cite{Harlander:2020duo} 
\begin{align}
Y_{FF}(t)
&=1+a_\mu \lp \frac{7}{8} C_A-\frac{\pi\beta_1}{\beta_0} \rp  \nonumber \\ 
&\quad{}+a_\mu^2 \,
\Bigg\{ \lp \frac{227}{180} -\frac{87}{80} \ln{2}+\frac{27}{40} \ln{3}
\rp C_A^2 \nonumber \\
&\qquad{}\qquad{}+
\lp -\frac{C_A}{18} +\frac{3}{16} C_F  \rp T_F n_f
-\frac{7 \pi\beta_1}{8\beta_0} C_A
+\frac{\pi^2\beta_1^2}{\beta_0^2}-\frac{\pi^2\beta_2}{\beta_0} \nonumber \\
&\qquad{}\qquad{}
+\left[ \frac{3}{32} C_A^2+\lp \frac{C_A}{8} +\frac{C_F}{4} \rp T_F n_f \right] L_z
\Bigg\}+\mathcal{O}(a_\mu^3) \, .
\end{align}


\section{All-order model for the perturbative Adler function and subtraction term}
\label{app:renormalonmodel}

In the main text, we employed a model for the Borel transform of the perturbative Adler function, originally proposed in ref.~\cite{Beneke:2008ad}, and the perturbative part of the action density that allowed us to visualize the working of the gradient-flow subtraction in higher orders than known by exact multi-loop computation results. We briefly review this model and its extension to the action density here, and refer to ref.~\cite{Beneke:2008ad} for details as well as definitions of expressions appearing in the equations below. 

The idea is  to merge the exactly known low-order coefficients with the asymptotic behaviour from the leading IR and UV renormalon singularities to model the series expansion and its Borel transform to all orders. For the Adler function, an estimate of the $\mathcal{O}(\alpha_s^5)$ 
coefficient $c_5=283$ is included \cite{Beneke:2008ad}, and then $c_1, \ldots c_5$ are employed to determine the five unknown parameters ($d_{0,1}^{\rm PO}$ and three Stokes constants) of the ansatz 
\begin{equation}
B[D](u) \,=\, B[\widehat{D}_1^{\rm UV}](u) + 
B[\widehat{D}_2^{\rm IR}](u) + B[\widehat{D}_3^{\rm IR}](u) +
d_0^{\rm PO} + d_1^{\rm PO} u 
\label{eq:modelD}
\end{equation}
for the Borel transform of the Adler function. The first 
three terms on the right-hand side incorporate the first 
UV renormalon and first two IR renormalon singularities with unknown 
Stokes constants, which together with $d_0^{\rm PO}$, $d_1^{\rm PO}$
are chosen such that  $c_1, \ldots c_5$ are reproduced exactly from the expansion of the Borel transform in $u$. Relative to ref.~\cite{Beneke:2008ad}, we now use the extension of eq.~(5.9) of ref.~\cite{Beneke:2008ad} to the $1/n^3$-correction, 
which uses the 5-loop beta-function coefficient to obtain $b_3$ in eq.~(5.4). Furthermore, for $\widehat{D}_2^{\rm IR}$, we now include the two-loop correction to the gluon-condensate short-distance coefficient $C_{FF}$, which completes the expression for the $1/n^2$ term.

For the gradient-flow subtraction term, $Y_1$, resp. the perturbative part of the action density in the small-flow time expansion, three low-order terms $e_1, e_2, e_3$ in the expansion of $E$ are available. However, the series expansion of $E$ has no UV renormalons, while the Stokes constant of the gluon condensate renormalon series $\widehat{E}_2^{\rm IR}$ is tied to  $D_2^{\rm IR}$ to effect the renormalon cancellation related to the universal standard gluon condensate. Hence no 
further information is required to determine the 
three remaining parameters of the ansatz
\begin{equation}
B[E](u) \,=\, B[\widehat{E}_2^{\rm IR}](u) + B[\widehat{E}_3^{\rm IR}](u) +
e_0^{\rm PO} + e_1^{\rm PO} u 
\label{eq:modelE}
\end{equation}
for the Borel transform of the series expansion of the subtraction term in terms of the three exactly known coefficients $e_1, e_2, e_3$. 
We employ eq.~(5.9) of ref.~\cite{Beneke:2008ad} with the above extensions accordingly.

To fully specify the ansatz, one needs to specify the renormalization scales $\mu_{\rm ref}^D$, resp. $\mu_{\rm ref}^{E}$, at which 
the known perturbative coefficients of the Adler function and the action density are evaluated. 
In this work, we choose $\mu_{\rm ref}^D=Q$ and 
$\mu_{\rm ref}^E=\sqrt{1/t}$, which results in the following numerical values of the parameters of the models eqs.~\eqref{eq:modelD}, \eqref{eq:modelE}:
\begin{equation}
\arraycolsep0.2cm
\begin{array}{llll}
     d_1^{\rm UV} = -0.0158979, 
    & d_2^{\rm IR} = 3.08572, 
    &d_3^{\rm IR} = -15.2998,  
    &
    \\[0.2cm]
   d_0^{\rm PO} = 18.2832,
    & d_1^{\rm PO} = -0.510789, 
    &\\[0.2cm]
    e_2^{\rm IR} = 0.468974, 
    & e_3^{\rm IR} = 2.84906, 
    & e_0^{\rm PO} = -2.76318, 
    & e_1^{\rm PO} = -0.0539007
\end{array}
\end{equation}
The conventions follow eq.~(6.2) in ref.~\cite{Beneke:2008ad}. 
We checked that varying $\mu_{\rm ref}^D$ and $\mu_{\rm ref}^E$, while changing the above parameters, always leads to the same qualitative behaviour of the higher-order series, which confirms the efficiency of gradient-flow subtraction. 


\section{Truncation of $r$ and uncertainty estimates}
\label{app:truncationorder}

In the main text we discuss the all-order perturbative series for 
$C_1(Q^2)$ and $Y_1(t)$ and the renormalon cancellation between them, but we truncate $r$, the ratio of gluon condensate short-distance coefficients, at next-to-next-to-leading order. 
One might wonder whether this compromises the 
perturbative accuracy of the main equation~\eqref{eq:newOPE}. 
In this appendix, we show that this is not the case and
that the truncation order of $C_1$ and $Y_1$ versus 
that of $r$ play different roles.

To clarify this issue, it is convenient to rewrite eq.~\eqref{eq:newOPE} in the form 
\be
D(Q^2)\propto C_1(Q^2)+\frac{r(Q^2,t)}{Q^4} \lp \frac{E(t)}{\pi^2}-\frac{Y_1(t)}{t^2}   \rp .
\label{eq:Drearranged}
\ee
Suppose also that $C_1$ and $Y_1$ are truncated 
at some order $N$, while $r$ is available through 
$\mathcal{O}(\alpha_s^k)$ with $k<N$. Since the expression in brackets,
\be
\frac{E(t)}{\pi^2}-\frac{Y_1(t)|_{\mathcal{O}(\alpha_s^{N})} }{t^2} 
=\mathcal{O}(\alpha_s^{N+1}/t^2,\LMS^4)\,,
\label{eq:EminusY1}
\ee
it follows that the two terms on the right-hand side of eq.~\eqref{eq:Drearranged} have errors\footnote{This should be considered as the perturbative truncation error. The error in the lattice computation of the non-perturbative action density $E(t)$ is not considered here.} 
\be
 \Delta D(Q^2) 
 = \mathcal{O}(\alpha_s^{N+1})+
 \mathcal{O}\!\left(\frac{\alpha_s^{N+1}}{t^2 Q^4}\right). 
\label{eq:neterror}
\ee
One sees that the parametric error is determined by
$N$ alone and employing a lower-order perturbative truncation perturbative for $r$ does not limit the accuracy of the result at the perturbative level.

Note that, while the $\mathcal{O}(\alpha_s^{N+1})$ errors originating from the truncated $C_1(Q^2)$ and $Y_1(t)$ are separately large 
when $N$ is in the regime where the $u=2$ renormalon is relevant,
the $\mathcal{O}(\alpha_s^{N+1})$ error in eq.~\eqref{eq:neterror} 
does not grow with a factorial coefficient, since the $u=2$ renormalon is largely cancelled between the two terms in eq.~\eqref{eq:neterror} by construction. In this regime around the minimal term of the series, the left-hand side of eq.~\eqref{eq:EminusY1} effectively counts as 
$\LMS^4$, since it approximates the standard gluon condensate. Thus, one sees that the truncation order $k$ of $r$ limits the non-perturbative accuracy of the gradient-flowed OPE including the gluon condensate to 
\begin{equation}
\Delta D(Q^2)  = \mathcal{O} \lp\alpha_s^{k+1} \LMS^4/Q^4 \rp,
\label{eq:D2}
\end{equation}
as expected. This also clarifies that the $u=1$ renormalon divergence of the series expansion of $r$, which leads to ambiguities of order $t\LMS^2$, $\LMS^2/Q^2$ in $r$, must be dealt with only when the gradient-flow subtracted OPE is extended to dimension-six operators, since eq.~\eqref{eq:D2} identifies this as a $\LMS^6$ effect for the Adler function. 

To summarize: the dominant truncation uncertainties of the gradient-flowed OPE when the dimension-four gluon condensate is included, are of order $\alpha_s^{N+1}$ (with no factorially large coefficients related to the $u=2$ renormalon), $\alpha_s^{N+k+2}/(t^2 Q^4)$ , $\alpha_s^{k+1} \LMS^4/Q^4$ and $t \LMS^6/Q^4$, $\LMS^6/Q^6$. 

\section{$t$ dependence of $\delta_{V+A}^{(0)}$}
\label{app:tdep}

\begin{figure}[t] 
\begin{center}
\includegraphics[width=11cm]{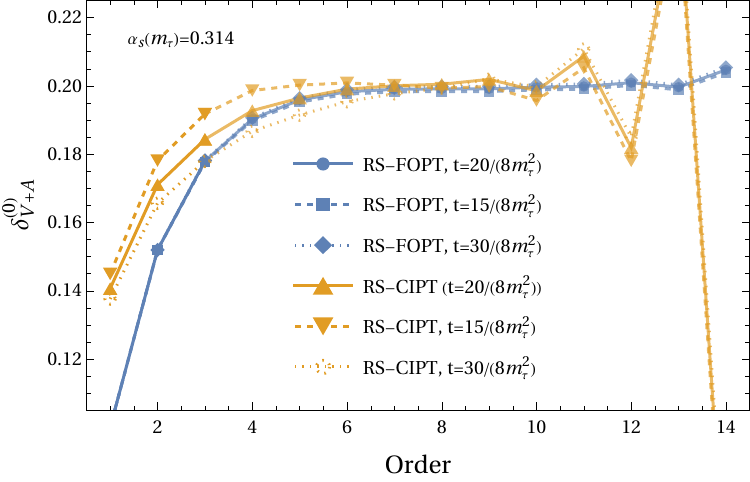}
\end{center}
\caption{
Comparison of the $t$ dependence of the subtracted FO (blue) and CI (orange) hadronic tau decay perturbative series. 
Beyond the third order, estimated perturbative coefficients are used, as indicated by fainter colours.}
\label{fig:FOvsCItdep} 
\end{figure}

In this appendix, we discuss the $t$-dependence of the gradient-flow subtracted hadronic tau decay series $\delta_{V+A}^{(0)}$ in both FOPT and CIPT. The numerical results are shown in 
fig.~\ref{fig:FOvsCItdep}. Since the effect of subtraction is strongly suppressed in FOPT, the $t$-dependence is also negligible. However, CIPT shows a strong dependence in intermediate orders. Nonetheless, all lines
shown converge to a certain region between the 6th to 10th order, regardless of the choice of $t$ or whether FO or CI is employed. In the following we explain these observations. 

Let us start by studying at which order $t$ dependence can enter. We express the perturbative series as
\be
C_1(Q^2)=1+\sum_{n \geq 0}  d_n(L) \alpha_s^{n+1} \,,
\ee
\be
Y_1(t)=\sum_{n \geq 0} e_n (L_t) \alpha_s^{n+1}\,,
\ee
\be
C_{FF}(Q^2)=\sum_{n \geq  0} h_n(L) \alpha_s^{n}\, ,
\ee
\be
Y_{FF}(t)=\sum_{n \geq  0} \ell_n(L_t) \alpha_s^{n}\,.
\ee
The perturbative contribution to $\delta_{\rm QCD}$ is given by
\begin{eqnarray}
\delta_{V+A}^{(0)}
&=&\frac{1}{2 \pi i} \oint_{|x|=1} \frac{dx}{x} 
\,W(x) 
\bigg[ \lp d_0-\frac{1}{m_{\tau}^4 t^2 x^2} \frac{e_0 h_0}{\ell_0} \rp \alpha_s\nonumber\\[0.2cm]
&&+\,\lp d_1(L)-\frac{1}{m_{\tau}^4 t^2 x^2} \frac{e_1(L_t) h_0 \ell_0+e_0 h_1 \ell_0-e_0 h_0 \ell_1}{\ell_0^2} \rp \alpha^2_s+\mathcal{O}(\alpha_s^3)
\bigg] \,,\qquad{}
\end{eqnarray}
where 
\be
W(x)=(1-x)^3 (1+x)\,,
\ee
and $\alpha_s$ without argument refers to the scale $\mu$.
Note that $L$ or $L_t$ dependence does not enter in some low-order coefficients and we omitted the argument for those coefficients.
$t \frac{d}{dt} \delta_{V+A}^{(0)}$ is given by
\begin{eqnarray}
t \frac{d}{dt} \delta^{(0)}_{V+A}
&=&\frac{1}{2 \pi i} \oint_{|x|=1} \frac{dx}{x} \,W(x) \,
\frac{1}{m_{\tau}^4 t^2 x^2} \,\bigg[ \, 2\,\frac{e_0 h_0}{\ell_0}  \alpha_s\nonumber\\[0.2cm]
&&
+ \, \lp 2 \,
 \frac{e_1 (L_t) h_0 \ell_0+e_0 h_1 \ell_0-e_0 h_0 \ell_1}{\ell_0^2} 
 -\frac{t \frac{\partial}{\partial t} e_1 (L_t) h_0 \ell_0}{\ell_0^2}
 \rp
\alpha^2_s
+\mathcal{O}(\alpha_s^3)
\bigg] .\qquad
\end{eqnarray}

In FOPT, $t \frac{d}{dt} \delta^{(0)}_{V+A}$ vanishes at $\mathcal{O}(\alpha_s^1)$ and $\mathcal{O}(\alpha_s^2)$, because the inclusive tau decay weight $W(x)$ does not contain an $x^2$ term and 
\be
\frac{1}{2 \pi i} \oint_{|x|=1} \frac{dx}{x} \,
W(x) \frac{1}{x^2}=0 .  \label{eq:xsqzero}
\ee
In FOPT, $t \frac{d}{dt} \delta_{V+A}^{(0)}$ becomes non-zero
only at $\mathcal{O}(\alpha_s^3)$,
where the integrand exhibits logarithmic dependence on $x$ through 
the dependence of $h_n(L)$ (with $n \geq 2$) on $L=\ln(-x m_\tau^2/\mu^2)$. In CIPT, however,  
$t \frac{d}{dt} \delta_{\rm QCD}^{\rm pert}$ is non-zero 
already at $\mathcal{O}(\alpha^1_s)$, because\footnote{For convenience, we use the square of the scale in the argument of $\alpha_s$ in this appendix.} 
\be
\frac{1}{2 \pi i} \oint_{|x|=1} \frac{dx}{x} \,
W(x) \frac{1}{x^2} \,\alpha_s(-m_{\tau}^2 x) \neq 0 \,.
\ee
This explains the observation that $t$ dependence is more significant in CIPT than in FOPT at low orders in perturbation theory.

There is an important difference in the $t$ dependence between FOPT and CIPT not only at low orders but at all orders. The difference can be easily understood in the large-$\beta_0$ approximation, which is therefore adopted in the following. We start from the general expression 
\be
\delta_{V+A}^{\text{large-$\beta_0$}}
=\sum_{n \geq 0} \frac{1}{2 \pi i} \oint_{|x|=1} 
\frac{dx}{x} \,W(x)  
\lp d_n(L)-\frac{r}{m_{\tau}^4 t^2 x^2} e_n(L_t) \rp \alpha_s^{n+1}.
\ee
In the large-$\beta_0$ approximation $r=h_0/l_0$ exactly and $r$ is an $L$-independent constant. The $t$-dependence is given by
\begin{align}
t \frac{d}{dt} \delta_{V+A}^{\text{large-$\beta_0$}}
&=\sum_{n \geq 0} \frac{1}{2 \pi i} \oint_{|x|=1} 
\frac{dx}{x} \,W(x)  \,
\frac{r}{m_{\tau}^4 t^2 x^2} \lp 2-t \frac{d}{dt} \rp e_n(L_t)  \alpha_s^{n+1} \non
&= \sum_{n \geq 0}  \frac{1}{2 \pi i} \oint_{|x|=1} 
\frac{dx}{x} \,W(x)\,
\frac{r}{m_{\tau}^4 t^2 x^2} \lp 2 e_n + \beta_0 n e_{n-1} \rp  \alpha_s^{n+1}\,.
\end{align}
We have used
\be
t \frac{d}{dt} e_n(L_t)
=-\frac{d}{d L_t} e_n(L_t)
=-\beta_0 n e_{n-1} ,
\ee
which follows from the RG equation.
The Borel transform of the 
perturbative series 
\be
\sum_{n \geq 0} (2 e_n+\beta_0 n e_{n-1})\, \alpha_s^{n+1}
\ee
is 
\be
B_{\text{$t$-dep}}(u)
=\sum_{n=0}^{\infty} \frac{2 e_n+\beta_0 n e_{n-1}}{n!} \lp \frac{u}{-\beta_0} \rp^{n}
=(2-u) B_E(u) \equiv  \lp 2 t \mu^2 e^{5/3}\rp^u  g(u)\,, 
\label{eq:Btdep}
\ee
where 
\be 
g(u) =  \frac{3}{4 \pi^3}\,\Gamma(3-u)
\label{eq:defgu}
\ee
follows from $B_E(u)$  defined in eq.~\eqref{eq:BorelE}. Note that 
the $u=2$ renormalon is eliminated in the $t$-dependence. 

Following the discussion of ref.~\cite{Hoang:2020mkw}, the all-order resummation of the series in terms of the Borel integral is given for FOPT by 
\be
t \frac{d}{dt} \delta_{V+A, \rm FOPT}^{\text{large-$\beta_0$}}
=-\frac{1}{\beta_0} \int_0^{\infty} \!\!du\,  \frac{1}{2 \pi i} \oint_{|x|=1} 
\frac{dx}{x} \,W(x)\,e^{\frac{u}{\beta_0 \alpha_s(-m_{\tau}^2 x)}}\,
\frac{r}{m_{\tau}^4 t^2 x^2} B_{\text{$t$-dep}}(u)|_{\mu^2=-m_{\tau}^2 x}  \,, \label{eq:tdepFOPT}
\ee
and for CIPT by
\begin{eqnarray}
&&t \frac{d}{dt} \delta_{V+A, \rm CIPT}^{\text{large-$\beta_0$}} =
-\frac{1}{\beta_0} \int_0^{\infty} \!\!du\,  
\nonumber\\
&& \hspace*{0.7cm}\times\,\frac{1}{2 \pi i} \oint_{\mathcal{C}_x} 
\frac{dx}{x} \,W(x)\,e^{\frac{u}{\beta_0 \alpha_s(m_{\tau}^2)}} \,
\frac{r}{m_{\tau}^4 t^2 x^2} \frac{\alpha_s(-m_{\tau}^2x)}{\alpha_s(m_{\tau}^2)} B_{\text{$t$-dep}}\!\lp \frac{\alpha_s(-m_{\tau}^2x)}{\alpha_s(m_{\tau}^2)} u \rp \!\bigg|_{\mu^2=-m_{\tau}^2 x}
\hspace*{-1cm}\, ,
\quad\qquad
\label{eq:tdepCIPT}
\end{eqnarray}
where $\mathcal{C}_x$ is a deformed contour that connects $x=1 \pm i0$
avoiding singularities in the 
$x$-plane, which is defined in ref.~\cite{Hoang:2020mkw}.
In FOPT, employing eq.~\eqref{eq:defgu}, the all-order result can be rewritten as
\begin{align}
t \frac{d}{dt} \delta_{V+A, \rm FOPT}^{\text{large-$\beta_0$}}
=-\frac{1}{\beta_0} \int_0^{\infty} \!\!du\,  \frac{1}{2 \pi i} \oint_{|x|=1} 
\frac{dx}{x}\, W(x)\,e^{\frac{u}{\beta_0 \alpha_s(1/(2te^{5/3}))}}\,
\frac{r}{m_{\tau}^4 t^2 x^2} \,g(u)\,.
\label{eq:FOPTtdep2}
\end{align}
This follows from  
\begin{align}
(-2 t m_{\tau}^2 x e^{5/3})^u \,e^{\frac{u}{\beta_0 \alpha_s(-m_{\tau}^2 x)}} =e^{\frac{u}{\beta_0} \left[ \beta_0 \log(-2 t m_{\tau}^2 x e^{5/3})+\frac{1}{\alpha_s(-m_{\tau}^2 x)} \right]} =e^{\frac{u}{\beta_0 \alpha_s(1/(2t e^{5/3}))}} \, , \label{eq:alphasrewrite}
\end{align}
where for the last equality the one-loop scale-dependence of $\alpha_s$ is used, which is exact in the large-$\beta_0$ approximation.
From eq.~\eqref{eq:FOPTtdep2}, where the $x$ dependence of the integrand is visible, it follows that $t \frac{d}{dt} \delta_{\rm QCD, FOPT}^{\text{large-$\beta_0$}}=0$, again due to eq.~\eqref{eq:xsqzero}. In fact, the perturbative contribution is already zero at every order in FOPT, since, as discussed earlier, the 
$t$ dependence is introduced through the $L$ dependence of $r$, which is not present in the large-$\beta_0$ approximation.

For the CIPT expression \eqref{eq:tdepCIPT} the same manipulations lead to
\begin{eqnarray}
&&t \frac{d}{dt} \delta_{V+A, \rm CIPT}^{\text{large-$\beta_0$}} = -\frac{1}{\beta_0} \int_0^{\infty} \!\!du\,  
\nonumber\\
&& \hspace*{0.7cm}\times\,\frac{1}{2 \pi i} \oint_{\mathcal{C}_x} \frac{dx}{x} \,W(x)\,
e^{\frac{u}{\beta_0 \alpha_s(1/t)} \frac{\alpha_s(- m_{\tau}^2x)}{\alpha_s(m_{\tau}^2)}}  \,
\frac{r}{m_{\tau}^4 t^2 x^2} \frac{\alpha_s(-m_{\tau}^2x)}{\alpha_s(m_{\tau}^2)}g\!\lp \frac{\alpha_s(-m_{\tau}^2x)}{\alpha_s(m_{\tau}^2)} u \rp  .
\qquad
\end{eqnarray}
A formal change of the integration variable 
$\frac{\alpha_s(-m_{\tau}^2x)}{\alpha_s(m_{\tau}^2)} u \to u$ 
and identification of the integration contour $\mathcal{C}_x$ with the circle $|x|=1$, renders the right-hand side  
exactly the same as eq.~\eqref{eq:FOPTtdep2}, which evaluated to zero.
However, this identification is not legitimate in the
presence of IR renormalons~\cite{Hoang:2020mkw}. While the $u=2$ renormalon has been cancelled in $g(u)$, those at $u=3$ and beyond remain.
We thus conclude that the remaining IR renormalon at $u=3$ 
and beyond causes a non-vanishing $t$-dependence even when the 
the CIPT series is summed to all orders, in contrast to the FOPT case, 
which is another manifestation that CIPT is not consistent with the analytic structure of the OPE.\footnote{We checked that the $t$ dependence of the CIPT series indeed does not 
go to zero at higher orders in the large-$\beta_0$ approximation.} 
From the confluence of CIPT lines in intermediate orders in fig.~\ref{fig:FOvsCItdep} one sees, however, that this effect is relatively small compared to the difference between the perturbative series themselves shown in figure~\ref{fig:FOvsCI}, since the former is caused only by the $u=3$ renormalon.

Finally, we make a remark. In the case of the subtracted Adler function,
the sizeable $t$ dependence was largely cancelled after adding the non-perturbative contribution to the perturbative one.
In the case of the hadronic decay width, in contrast, $t$ dependence was largely diminished within the perturbative contribution, as seen in fig.~\ref{fig:FOvsCItdep}. The reason has been clarified by the above argument. However, the present conclusion is specific to the total hadronic decay width, for which the weight function $W(x)$ has no $x^2$ term and eq.~\eqref{eq:xsqzero} holds, but it does not hold for general spectral weights.\footnote{See ref.~\cite{Beneke:2012vb} for the comparison of FOPT vs. CIPT behaviour for different spectral weights.} In these more general situations, we expect similar behaviour as for the Adler function. Yet, 
the above discussion clearly highlights a difference between 
FOPT and CIPT.

\bibliography{refs}

\end{document}